\documentclass[twocolumn,aps,prd]{revtex4}
\usepackage{calc}
\newcommand{\ra}{\rangle}
\newcommand{\la}{\langle}

\newcommand{\ov}{\overline}
\newcommand{\cd}{\! \cdot \!}
\newcommand{\be}{\begin{equation}}
\newcommand{\ee}{\end{equation}}
\newcommand{\ba}{\begin{eqnarray}}
\newcommand{\ea}{\end{eqnarray}}
\renewcommand{\slash}{ \not}
\newcommand{\fermionright}[2]{ \put(#1,#2){\vector(1,0){13}}
\put(#1,#2){\line(1,0){20}} }
\newcommand{\fermionleft}[2]{ \put(#1,#2){\vector(-1,0){13}}
\put(#1,#2){\line(-1,0){20}} }
\begin{document}
\title{Quark Schwinger-Dyson Evaluation of the l1, l2 Coefficients
in the Chiral Lagrangian.}

\author{F. J. Llanes-Estrada } \email{fllanes@fis.ucm.es}
\affiliation{Depto. de F\'{\i}sica Te\'orica I,  Univ.
Complutense. 28040 Madrid, Spain. }
\author{P. de A. Bicudo}  \email{bicudo@ist.utl.pt}
\affiliation{CFIF, Instituto Superior T\'ecnico, Avda. Rovisco Pais
49001 Lisboa, Portugal}

\begin{abstract}
 Using a systematic expansion of the quark-antiquark
Bethe-Salpeter wavefunctions in the relativistic quark model and working 
to $O(P^4)$, in the chiral limit,
we are able to derive theoretical expressions relating the coefficients of
the chiral lagrangian $l_1$, $l_2$ to the underlying quark-antiquark
wavefunctions and interaction kernels. This is accomplished by using a
novel technique based on a Ward Identity for the quark-antiquark ladder
kernel which greatly simplifies the required effort. 
Numerical evaluations are performed in two simple specific models.
\end{abstract}
\maketitle

%%%%%%%%%%%%%%%%%%%%%%%%%%%%%%%%%%%%%%%%%%%%%%%%%%%%%%%%%%%%%%%%%%%
\section{Introduction.}\label{intro}
%%%%%%%%%%%%%%%%%%%%%%%%%%%%%%%%%%%%%%%%%%%%%%%%%%%%%%%%%%%%%%%%%%%
It has traditionally been considered a triumph of theoretical physics
when the parameters of an effective, low energy theory which correctly
describe phenomena at a given scale can be related to those of an
underlying, more fundamental scheme of thought which grounds it. Brilliant
examples are Fritz London's explanation of the quantum nature of the Van
der Waals forces \cite{London} or the derivation of the atomic
relativistic corrections as a consequence of the Dirac's equation for the
electron.
Low energy hadronic processes are interpreted with the aid of mainly two
types of theories: nucleon-nucleon non relativistic interactions such as
the Nijmegen \cite{Nijmegen} or Argonne \cite{Argonne} potentials, for the
heavier hadrons, and relativistic chiral lagrangians \cite{Gasser} for the
lightest components, the pions.

The deeper quark theories such as QCD or any microscopic models thereof
pretend in principle to describe the totality of hadronic physics. They attempt
to be complete descriptions of hadronic processes. Unfortunately, the
complexity of many body hadronic calculations makes it forbidding  to
fully exploit the underlying scheme, and maintain the validity of the
low energy effective theory.

As a consequence, an initial goal for the microscopic theory should be to
reproduce in some limit the macroscopic models and to relate their
parameters to its own set (hopefully smaller).
In this paper we make the case for microscopic quark models inspired in
QCD as generating the parameters of the chiral lagrangian. This
lagrangian, describing the low energy behavior of a pion system, and
being able to incorporate the coupling of pions to other mesons (as much
as the low energy theorems of PCAC \cite{Renner} do) is universal (in
the sense that any theory with the same symmetries can be cast in its
form) and provides a consistent derivative expansion in powers of the
momentum and mass of any pions present in a system, divided by a typical
scale of the strong interactions.

Unfortunately, this derivative expansion has to incorporate new
coefficients order by order. These new coefficients absorb the divergences
of loops generated by the vertices of smaller order terms, and so their
value is generally renormalized. Still, the common usage of this
Lagrangian \cite{Pelaez} proceeds by fitting this coefficients to some
observable set at a given scale.

We show how these coefficients can be related systematically to quark level
parameters in the planar approximation. This has been accomplished in the 
past for the simplest,
$O(P^2)$ chiral lagrangian whose parameters are only two, in usual
notation, $M_\pi$, $f_\pi$ (the pion mass and decay constant). To this
order, these two parameters are conventionally set to take their physical
value.
To the next order, the lagrangian contains six parameters,
which generate the $O(P^4)$ vertices, $l_1$, $l_2$, absorbing
divergences in the 4 pion Green function, $l_3$, $l_4$ which absorb
counterterms of the mass and axial current renormalizations, and finally
$M_\pi$, $f_\pi$. The complete renormalization scheme is specified in
\cite{Gasser}.
The parameters $M_\pi$, $f_\pi$ have long been accounted for by relativistic quark
models \cite{Maris,Barcelona}. The $l$'s on the other hand have not 
been treated in quark models with non-contact interactions.
We insist in the point that any theory which respects the $SU(2)_L  \times
SU(2)_R$ chiral symmetry breaking pattern, let it be a Nambu-Jona-Lasinio
quark theory \cite{NJL}, a large $N_c$ 
expansion \cite{andrianov}, a string theory or any other exotic creation, 
can be cast in
the form of the chiral lagrangian  and the only difference between all
of them is the numerical values of the $l_i$ coefficients.

It is therefore of paramount importance to determine them from the
theories which we believe correctly describe the physics at the GeV scale,
in terms of quarks and antiquarks. Lattice determinations are making
progress in that direction \cite{Nelson}, but the Schwinger-Dyson equation
formalism should provide an alternative determination in the near future.
An interesting paper \cite{Roberts} exists where, at the lagrangian level,
the action for a relativistic quark model is bosonized to obtain an
effective meson lagrangian, used then to calculate pion-pion scattering
lengths. We are going to theoretically extend this approach in two
directions.
First, we will start with the most general chirally symmetric quark model,
in which the pion is well described by a quark-antiquark pair after chiral
symmetry breaking (encompassing in this way an ample spectrum of models)
and by using their chiral properties, reduce the four pion Green's
functions to a minimal set of diagrams. In this way, no bosonization is
performed, and at all steps the way quark interactions arrange themselves
to comply with the chiral theorems is explicitly visible.
Second, comparing the result with the same
calculation in a chiral lagrangian formalism, one can immediately read
off the $l_i$ coefficients of the chiral lagrangian in terms of diagrams
which can numerically be calculated in the quark model. This
rather technical numerical evaluation will be simplified
in this work by confining ourselves to simple, finite models, although
the numerical results will then be limited.
The method used here has already successfully being exploited to
demonstrate how this class of models comply with the Weinberg theorem in
\cite{weall,Pedrosolo,CotanchMaris}. The Weinberg theorem
was derived with an expansion to $O(P^2)$,
$O({M_\pi}^2)$. We now concentrate on the  $O(P^4)$, $O({M_\pi}^0)$ chiral
lagrangian, that is, the only parameters are $f_\pi$, $l_1$, and $l_2$. We
will perform the same expansion in the quark-antiquark diagrams and
compare the results to read off $l_1$, $l_2$. The expansion will be
carried out whenever possible in a Feynman diagram language to avoid
lengthy expressions for the sake of readability.
The rest of this paper is organized as follows:
in section \ref{ChiLag} we briefly settle the notation for our chiral
perturbation theory discussion and remind the reader of a few well-known
facts in this field. Section \ref{WI} settles the notation of the
microscopic quark manipulations to follow and provides the reader with a
useful chiral Ward Identity recently introduced \cite{weall},
\cite{Pedrosolo}. Section \ref{pionpion} is the core of the paper and
presents the reduction of the pion scattering amplitude, whereas the
resulting diagrams are calculated in two simple models in section
\ref{modelitos}. Some issues clarifying the normalization of the 
Bethe-Salpeter equation are relegated to the appendix.

%%%%%%%%%%%%%%%%%%%%%%%%%%%%%%%%%%%%%%%%%%%%%%%%%%%%%%%%%%%%%%%%%%%
\section{Chiral Lagrangian of order $P^4$.} \label{ChiLag}
%%%%%%%%%%%%%%%%%%%%%%%%%%%%%%%%%%%%%%%%%%%%%%%%%%%%%%%%%%%%%%%%%%%
The macroscopic theory one generally writes down for pion fields
alone is to lowest order the Non
Linear Sigma Model. One can proceed by constructing, from the three pion
fields, $\vec{\pi}=(\pi_1,\pi_2,\pi_3)$, a 4-vector normalized to one
(this normalization is equivalent to eliminating the explicit $\sigma$
degree of freedom from the Linear Sigma Model)
\begin{equation}
U= \left[
\begin{array}{c}
\sqrt{1-\frac{\vec{\pi}^2}{{F}^2}} \\
\frac{\vec{\pi}}{F}
\end{array}
\right]
\end{equation}
and then constructing Lorentz scalar, parity invariant terms.
To $O(P^4)$ that lagrangian can be extended by terms which in the chiral
limit ($m_q=0$) have to be of the form \cite{Gasser}
\begin{equation} \label{chirallag}
{\mathcal L}^{(4)} = \frac{1}{{F}^4} ( l_1 (\vec{\pi}_{, \mu} \cd
\vec{\pi}^{, \mu} )(\vec{\pi}_{, \nu} \cd \vec{\pi}^{, \nu})+
l_2 (\vec{\pi}^{, \mu} \cd \vec{\pi}^{, \nu} )(\vec{\pi}_{, \mu} \cd
\vec{\pi}_{, \nu} )  )
\end{equation}
where the scalar product dots are in isospin space.
This lagrangian is on shell, for massless pions (else the $l_3$,
$l_4$ counterterms should also be present) and contributes at tree level
to the $O(P^4)$
pion-pion scattering amplitude, and it is this contribution which we aim
to reproduce microscopically. In the chiral formalism, there are also
one-loop contributions from the $O(P^2)$ lagrangian which we do not
consider in this work, since our quark-level calculation will not be
extended to meson loops. Therefore, to this level, it is fair to compare
our results only with those obtained in chiral perturbation theory without
meson loops.
With this caveat in mind, the pion-pion scattering amplitudes are
straightforwardly obtained. By using crossing symmetry, the different
isospin channels can be related in terms of only one amplitude $A$:
\begin{eqnarray}
T_{I=2}=A(t,s,u)+A(u,t,s) \\ \nonumber
T_{I=1}=A(t,s,u)-A(u,t,s) \\ \nonumber
T_{I=0}=3 A(s,t,u)+A(t,s,u)+A(u,t,s) \ .
\end{eqnarray}
This amplitude $A(s,t,u)$ can be obtained from the process $\pi_+ 
\pi_- \longrightarrow \pi_0 \pi_0$. Due to the final state Bose 
symmetry, and in the chiral limit when the Mandelstam variables 
satisfy $s+t+u=0$, the most general amplitude of order $P^4$ 
containing the polynomials $s^2$, $t^2$, $u^2$, $st$, $su$, $tu$, 
reduces to $A_1 s^2 + A_2 (t-u)^2$. The coefficients obtained from
the lagrangian (\ref{chirallag}) above yield 
\be \label{standardamplitude}
A^{(4)}(s,t,u)=\frac{1}{{F}^4} [(2 l_1 + \frac{l_2}{2})s^2 + 
\frac{l_2}{2} (t-u)^2] 
\ . \ee 

A full discussion of this and related issues, for example the relation 
between $f_\pi$ and $F$ which we further ignore in this paper to the
order we are working can be found in \cite{Gasser, Lehman}.

%%%%%%%%%%%%%%%%%%%%%%%%%%%%%%%%%%%%%%%%%%%%%%%%%%%%%%%%%%%%%%%%%%
\section{Notation for Quark Models and Chiral Ward Identities.}
\label{WI}
%%%%%%%%%%%%%%%%%%%%%%%%%%%%%%%%%%%%%%%%%%%%%%%%%%%%%%%%%%%%%%%%%%
A pion with momentum $P$ couples in relativistic models to fermion lines
whose momenta will be denoted by $k$, $k'$. In the massless quark limit, 
whenever $P=0$, then $k=k'$.
 We start by considering the bare
fermion propagator from any standard quark theory,
\be \label{barepropagator}
S_0(k)=\frac{i}{\slash k - m +i \epsilon}
\ee
and, after spontaneous chiral symmetry breaking mediated by a strong
interaction \cite{BicudoRibeiro}, \cite{Orsay}, the full fermion
propagator parameterized as
\be
S(k)=\frac{i}{A(k^2) \slash k - B(k^2) +i \epsilon}
\ee
which we take to be a solution of the planar Rainbow Schwinger-Dyson equation
\be \label{SD}
S(k)^{-1}=S_0(k)^{-1} - \int \frac{d^4q}{(2\pi)^4} V^a S(k+q) V_a K(q) \
.
\ee
We further define the bare axial vertex which couples a quark-antiquark
pair to a pseudoscalar current by means of the shorthand $\gamma_A^a$ (notice
that $m_u=0=m_d$ in the chiral limit employed in this paper):
\be
\gamma_A^a= \frac{\sigma^a}{2}(-i P_\mu \gamma^\mu \gamma_5 + 2im_u \gamma_5)
\ee
which satisfies
\be \label{smallGAvertex}
\gamma_A^a(k,k')=\frac{\sigma^a}{2} (S_0^{-1}(k) \gamma_5 + 
\gamma_5 S_0^{-1}(k'))
\ee
and the dressed axial vertex, dressed with a planar ladder $\Gamma_A$,
given by
\be \label{GAvertex}
\Gamma_A^a(k,k') = \gamma_A^a(k,k') + \int V^a S(k_1+q) \Gamma_A^a
S(k_2+q) V_a K(q)
\ee
or reconstructing the planar ladder expansion (in graphical form):
\begin{equation} \label{ladderDEF}
%>>>>>>>>>>>>>>>>>>>>>>>>>>>>>>>>>>>>>>>>>>>>>>>>>>>>>>>>>>>>>>
\begin{picture}(232,30)(0,0) %(63,761)               %791-Y
%\put(0,0){\pframebox(232,30){}}
\put(0,10){                         %inicio equacao
%%%%%%%%%%%%%%%%%%%%%%%%%%%%% ladder
\begin{picture}(40,20)(0,7)
\put(0,15){\line(1,0){5}}
\put(0,5){\vector(1,0){10}}
\put(15,15){\vector(-1,0){10}}
\put(15,5){\line(-1,0){5}}
\put(15,0){\framebox(10,20){}}
\put(25,15){\line(1,0){5}}
\put(25,5){\vector(1,0){10}}
\put(40,15){\vector(-1,0){10}}
\put(40,5){\line(-1,0){5}}
\end{picture}
$ \ = \ $
\begin{picture}(15,20)(0,7)
\put(0,15){\line(1,0){5}}
\put(0,5){\vector(1,0){10}}
\put(15,15){\vector(-1,0){10}}
\put(15,5){\line(-1,0){5}}
\end{picture}
$ \ + \ $
%%%%%%%%%%%%%%%%%%%%%%%%%% potential
\begin{picture}(30,20)(0,7)
\put(0,0){
\begin{picture}(15,20)(0,0)
\put(0,15){\line(1,0){5}}
\put(0,5){\vector(1,0){10}}
\put(15,15){\vector(-1,0){10}}
\put(15,5){\line(-1,0){5}}
\end{picture}}
\put(15,0){
\begin{picture}(15,20)(0,0)
\multiput(0,3)(0,2){6}{$\cdot$}
\put(0,15){\line(1,0){5}}
\put(0,5){\vector(1,0){10}}
\put(15,15){\vector(-1,0){10}}
\put(15,5){\line(-1,0){5}}
\end{picture}}
\end{picture}
$ \ + \ $
\begin{picture}(45,20)(0,7)
\put(0,0){
\begin{picture}(15,20)(0,0)
\put(0,15){\line(1,0){5}}
\put(0,5){\vector(1,0){10}}
\put(15,15){\vector(-1,0){10}}
\put(15,5){\line(-1,0){5}}
\end{picture}}
\put(15,0){
\begin{picture}(15,20)(0,0)
\multiput(0,3)(0,2){6}{$\cdot$}
\put(0,15){\line(1,0){5}}
\put(0,5){\vector(1,0){10}}
\put(15,15){\vector(-1,0){10}}
\put(15,5){\line(-1,0){5}}
\end{picture}}
\put(30,0){
\begin{picture}(15,20)(0,0)
\multiput(0,3)(0,2){6}{$\cdot$}
\put(0,15){\line(1,0){5}}
\put(0,5){\vector(1,0){10}}
\put(15,15){\vector(-1,0){10}}
\put(15,5){\line(-1,0){5}}
\end{picture}}
\end{picture}
$ \ + \dots  $
}
\end{picture}
%<<<<<<<<<<<<<<<<<<<<<<<<<<<<<<<<<<<<<<<<<<<<<<<<<<<<<<<
\end{equation}

(\ref{GAvertex}) takes the form
\begin{equation} \label{GAvertexgraph}
%>>>>>>>>>>>>>>>>>>>>>>>>>>>>>>>>>>>>>>>>>>>>>>>>>>
\begin{picture}(180,25)(0,0)
%\put(0,0){\framebox(180,25){}}
\put(20,10){$S \, \Gamma_{A}^a \,S \, \,=$}
\put(65,10){
%%%%%%%%%%%%%%%%%%%%%%%%%%%%%% half closed box
\begin{picture}(100,20)(0,7)
\put(0,15){\line(1,0){5}}
\put(0,5){\vector(1,0){10}}
\put(15,15){\vector(-1,0){10}}
\put(15,5){\line(-1,0){5}}
\put(15,0){\framebox(10,20){}}
\put(25,15){\line(1,0){5}}
\put(25,5){\vector(1,0){10}}
\put(40,15){\vector(-1,0){10}}
\put(40,5){\line(-1,0){5}}
\put(40,10){\oval(10,10)[r]}
\put(43,8){$\bullet$}
\put(49,10){$\gamma_{\hspace{-.08cm}A}^a$}
\end{picture}
}
\end{picture}
%<<<<<<<<<<<<<<<<<<<<<<<<<<<<<<<<<<<<<<<<<<<<<
\end{equation}
from which one can deduce the axial vector Ward Identity:
\be \label{GAvertex2}
\Gamma_A^a(k,k')=\frac{\sigma^a}{2}(S^{-1}(k) \gamma_5 + 
\gamma_5 S^{-1}(k')) := \frac{\sigma^a}{2}\Gamma_A \ .
\ee
This is analogous to the abelian vector Ward-Takahashi Identity
which in terms of the vertex $\Gamma_\mu$ defined by 
\begin{equation} \label{Vectovertex}
%>>>>>>>>>>>>>>>>>>>>>>>>>>>>>>>>>>>>>>>>>>>>>>>>>>
\begin{picture}(180,25)(0,0)
%\put(0,0){\framebox(180,25){}}
\put(20,10){$S \, \Gamma_\mu^a \,S \, \,=$}
\put(65,10){
%%%%%%%%%%%%%%%%%%%%%%%%%%%%%% half closed box
\begin{picture}(100,20)(0,7)
\put(0,15){\line(1,0){5}}
\put(0,5){\vector(1,0){10}}
\put(15,15){\vector(-1,0){10}}
\put(15,5){\line(-1,0){5}}
\put(15,0){\framebox(10,20){}}
\put(25,15){\line(1,0){5}}
\put(25,5){\vector(1,0){10}}
\put(40,15){\vector(-1,0){10}}
\put(40,5){\line(-1,0){5}}
\put(40,10){\oval(10,10)[r]}
\put(43,8){$\bullet$}
\put(49,10){$\gamma_\mu$}
\end{picture}
}
\end{picture}
%<<<<<<<<<<<<<<<<<<<<<<<<<<<<<<<<<<<<<<<<<<<<<
\end{equation}
yields
\be \label{VWI}
i (k_\mu-k'_\mu) \Gamma_\mu(k,k')= S^{-1}(k')-S^{-1}(k) \ .
\ee
Next we introduce the bound state formalism for quark-antiquark systems.
 To this end we remind the reader of the Bethe-Salpeter (BS) amplitude
$\chi$ (see
\cite{Maris}, \cite{Roberts}) for further details) which satisfies a
homogeneous Bethe-Salpeter equation:
\be \label{BS}
\chi^b(P,k)=\int V_a S(k'\! +\! \frac{P}{2}) \chi^b(P,k') S(k' \!
- \! \frac{P}{2}) V^a K(k-k') \ .
\ee
or in graphical form:
\be
\begin{picture}(215,40)(0,0)
%\put(0,0){\framebox(215,40){}}
\fermionright{5}{10}
\fermionleft{25}{30}
\put(10,4){$_k$} \put(8,34){$_{k  - \! P}$}
\put(25,10){\line(1,1){10}}  \put(25,30){\line(1,-1){10}}
{\linethickness{2pt} \put(35,20){\line(1,0){10}}}
\fermionright{105}{10}
\fermionleft{125}{30}
\put(110,4){$_k$} \put(108,34){$_{k -\! P}$}
\put(50,18){$\chi_\pi^b(P,k) =$}
\put(170,18){$\chi_\pi^b(P,k')$}
\multiput(125,8)(0,2){10}{$\cdot$}
\fermionright{125}{10}
\fermionleft{145}{30}
\put(130,4){$_{k'}$} \put(128,34){$_{k'-P}$}
\put(145,10){\line(1,1){10}} \put(145,30){\line(1,-1){10}}
{\linethickness{2pt} \put(155,20){\line(1,0){10}}   }
\end{picture}
\ee

Each incoming or outgoing pion in a particular process must contribute
with one of these $\chi$ functions, which carry pseudoscalar quantum
numbers by construction \cite{Liu}. The BS amplitude for a
particular pion depends on the total momentum of the pion $P$, and the
momentum of its fermion component $k \pm P/2$. Notice that this equation is the
homogeneous part of (\ref{GAvertex}) above when we interpret the pion
momentum $P$ as $k-k'$ in the vertex definition.

Now let us deepen our study of the vertex $\Gamma_A$. From
(\ref{GAvertex2})
can easily be seen that in the chiral limit ($m_q$=0),
\be \label{GammaB}
\Gamma_A^a(k,k'=k)=2i B(k^2) \gamma_5 \frac{\sigma^a}{2}\ ,
\ee
in terms of the SD amplitude $B$ solution of (\ref{SD}).
Equations (\ref{BS}) and (\ref{GAvertex}), homogeneous and not
homogeneous,
coincide when
$\gamma_A=0$. This is satisfied in the limit $m_q=0$ when also $P=0$ as
can be seen explicitly from (\ref{smallGAvertex}) and allows us to
identify, up to a normalization constant, $\chi_\pi (P=0,k)$ with
$\Gamma_A(k,k'=k)$. This constant coincides with $i f_\pi$, the pion decay
constant in the chiral limit (the proof is sketched in appendix 
\ref{normissues})
 and finally entails,
in combination with (\ref{GammaB}),
\be
\chi_\pi^a(P=0,k)=\frac{-i\Gamma^a_A(k,k'=k)}{f_\pi}=\frac{2 B(k^2)}{f_\pi}
\gamma_5 \frac{\sigma^a}{2}\ .
\ee
In \cite{Scadron} the proof was given why this BS amplitude, in
connection with the Axial Vector Ward Identity makes the pion a Goldstone
boson. In terms of our notation this has been rewritten in \cite{Pedrosolo}.

This discussion suggests a strategy
to systematically organize the corrections to the chiral, low momentum
limit, in an analogous fashion to that used in Chiral Perturbation Theory.
Since the vertex $\Gamma_A$ and diagrams constructed thereof satisfy
interesting Chiral identities, let us define
\be \label{expandGamma}
\chi^a(P,k) = \frac{-i \Gamma_A^a(P,k) + \Delta^a(P,k)}{f_\pi}
\ee
where the function $\Delta(P,k)$ so introduced can be expanded in a Taylor
series for low $P$. This expansion will organize the momentum corrections
to any diagram. One will first use the chiral results for $\Gamma_A$,
which will provide one with exact low energy theorems, and the numerical
corrections as $P$ is increased can then be expressed as overlaps of
$\Delta$ functions.

We do not yet specify the color, spin, flavor or momentum structure
of the interaction kernel and vertices $V_aV^aK(q)$, except
for one property: it must be chiral symmetry preserving, that is, V
commutes with $\gamma_5$. This guarantees the satisfaction of the
following Chiral Ward Identity (also discussed in \cite{weall},
\cite{Pedrosolo}) which proved essential:
\begin{eqnarray} \label{upperWI}
%>>>>>>>>>>>>>>>>>>>>>>>>>>>>>>>>>>>>>>>>>>>>>>
\begin{picture}(85,40)(0,0)
\put(5,10){
%%%%%%%%%%%%%%%%%%%%%%%%%%%%%% box with two gamma_{\hspace{-.08cm}A}s
\begin{picture}(40,20)(0,0)
\put(0,15){\line(1,0){5}}
\put(0,5){\vector(1,0){10}}
\put(15,15){\vector(-1,0){10}}
\put(15,5){\line(-1,0){5}}
\put(15,0){\framebox(10,20){}}
\put(25,15){\line(1,0){5}}
\put(25,5){\vector(1,0){10}}
\put(40,15){\vector(-1,0){10}}
\put(40,5){\line(-1,0){5}}
\put(40,10){\oval(30,10)[r]}
\put(42,2){$\bullet$}
\put(48,-6){$\Gamma_{\hspace{-.08cm}A}$}
\put(42,13){$\bullet$}
\put(48,18){$\Gamma_{\hspace{-.08cm}A}$}
\put(55,10){\vector(0,1){2}}
\end{picture}
}
\put(75,17){=}
\end{picture}
%<<<<<<<<<<<<<<<<<<<<<<<<<<<<<<<<<<<<<<<<<<<<<<<<<<<
%>>>>>>>>>>>>>>>>>>>>>>>>>>>>>>>>>>>>>>>>>>>>>>>>>
\begin{picture}(120,40)(0,0)
%%%%%%%%%%%%%%%%%%%%%%%%%%%%%% box
\put(5,10){\begin{picture}(40,20)(0,0)
\put(0,15){\line(1,0){5}}
\put(0,5){\vector(1,0){10}}
\put(15,15){\vector(-1,0){10}}
\put(15,5){\line(-1,0){5}}
\put(15,0){\framebox(10,20){}}
\put(25,15){\line(1,0){5}}
\put(25,5){\vector(1,0){10}}
\put(40,15){\vector(-1,0){10}}
\put(40,5){\line(-1,0){5}}
\end{picture}}
%%%%%%%%%%%%%%%%% intermediate Gamma_{\hspace{-.08cm}A}
\put(5,10){\begin{picture}(60,20)(0,0)
\put(38,2){$\bullet$}
\put(35,-8){$S^{-1}$}
\put(38,13){$\bullet$}
\put(37,20){$\Gamma_{\hspace{-.08cm}A}$}
\end{picture}}
%
%%%%%%%%%%%%%%%%%%%%%%%%%%%%%% box
\put(45,10){\begin{picture}(40,20)(0,0)
\put(0,15){\line(1,0){5}}
\put(0,5){\vector(1,0){10}}
\put(15,15){\vector(-1,0){10}}
\put(15,5){\line(-1,0){5}}
\put(15,0){\framebox(10,20){}}
\put(25,15){\line(1,0){5}}
\put(25,5){\vector(1,0){10}}
\put(40,15){\vector(-1,0){10}}
\put(40,5){\line(-1,0){5}}
\end{picture}}
%%%%%%%%%%%%%%%%%%%%%%%%%%%%%% half closed
\put(45,10){\begin{picture}(40,20)(0,0)
\put(40,10){\oval(10,10)[r]}
\put(43,8){$\bullet$}
\put(49,10){$\gamma_{\hspace{-.08cm}A}$}
\end{picture}}
\put(105,18){=}
\end{picture}
%<<<<<<<<<<<<<<<<<<<<<<<<<<<<<<<<<<<<<<<<<<<<<<<
\\ \nonumber
%>>>>>>>>>>>>>>>>>>>>>>>>>>>>>>>>>>>>>>>>>>>>>>>>>
\begin{picture}(105,30)(0,0)
%%%%%%%%%%%%%%%%%%%%%%%%%%%%%% box
\put(15,5){\begin{picture}(40,20)(0,0)
\put(0,15){\line(1,0){5}}
\put(0,5){\vector(1,0){10}}
\put(15,15){\vector(-1,0){10}}
\put(15,5){\line(-1,0){5}}
\put(15,0){\framebox(10,20){}}
\put(25,15){\line(1,0){5}}
\put(25,5){\vector(1,0){10}}
\put(40,15){\vector(-1,0){10}}
\put(40,5){\line(-1,0){5}}
\end{picture}}
%%%%%%%%%%%%%%%%%%%%%%%%%%%%%% half closed
\put(15,5){\begin{picture}(40,20)(0,0)
\put(40,10){\oval(10,10)[r]}
\put(43,8){$\bullet$}
\put(-13,15){$\gamma_5$}
\put(49,10){$\gamma_{\hspace{-.08cm}A}$}
\end{picture}}
\put(90,14){+}
\end{picture}
%<<<<<<<<<<<<<<<<<<<<<<<<<<<<<<<<<<<<<<<<<<<<<<
%>>>>>>>>>>>>>>>>>>>>>>>>>>>>>>>>>>>>>>>
\begin{picture}(75,30)(0,0)
%%%%%%%%%%%%%%%%%%%%%%%%%%%%%% box
\put(5,5){\begin{picture}(40,20)(0,0)
\put(0,15){\line(1,0){5}}
\put(0,5){\vector(1,0){10}}
\put(15,15){\vector(-1,0){10}}
\put(15,5){\line(-1,0){5}}
\put(15,0){\framebox(10,20){}}
\put(25,15){\line(1,0){5}}
\put(25,5){\vector(1,0){10}}
\put(40,15){\vector(-1,0){10}}
\put(40,5){\line(-1,0){5}}
\end{picture}}
%%%%%%%%%%%%%%%%%%%%%%%%%%%%%% half closed
\put(5,5){\begin{picture}(40,20)(0,0)
\put(40,10){\oval(10,10)[r]}
\put(43,8){$\bullet$}
\put(49,10){$\gamma_5 \gamma_{\hspace{-.08cm}A}$}
\end{picture}}
\end{picture}
%<<<<<<<<<<<<<<<<<<<<<<<<<<<<<<<<<<<<<<<<<<<<<<<<<<<<
\ .
\end{eqnarray}
or, for a general vertex not necessarily pseudoscalar:

\begin{eqnarray} \label{upperWI2}
%>>>>>>>>>>>>>>>>>>>>>>>>>>>>>>>>>>>>>>>>>>>>>>>>>
\begin{picture}(120,40)(0,0)
%%%%%%%%%%%%%%%%%%%%%%%%%%%%%% box
\put(5,10){\begin{picture}(40,20)(0,0)
\put(0,15){\line(1,0){5}}
\put(0,5){\vector(1,0){10}}
\put(15,15){\vector(-1,0){10}}
\put(15,5){\line(-1,0){5}}
\put(15,0){\framebox(10,20){}}
\put(25,15){\line(1,0){5}}
\put(25,5){\vector(1,0){10}}
\put(40,15){\vector(-1,0){10}}
\put(40,5){\line(-1,0){5}}
\end{picture}}
%%%%%%%%%%%%%%%%% intermediate Gamma_{\hspace{-.08cm}A}
\put(5,10){\begin{picture}(60,20)(0,0)
\put(38,2){$\bullet$}
\put(35,-8){$S^{-1}$}
\put(38,13){$\bullet$}
\put(37,20){$\Gamma_{\hspace{-.08cm}A}$}
\end{picture}}
%
%%%%%%%%%%%%%%%%%%%%%%%%%%%%%% box
\put(45,10){\begin{picture}(40,20)(0,0)
\put(0,15){\line(1,0){5}}
\put(0,5){\vector(1,0){10}}
\put(15,15){\vector(-1,0){10}}
\put(15,5){\line(-1,0){5}}
\put(15,0){\framebox(10,20){}}
\put(25,15){\line(1,0){5}}
\put(25,5){\vector(1,0){10}}
\put(40,15){\vector(-1,0){10}}
\put(40,5){\line(-1,0){5}}
\end{picture}}
%%%%%%%%%%%%%%%%%%%%%%%%%%%%%% half closed
\put(45,10){\begin{picture}(40,20)(0,0)
\put(40,10){\oval(10,10)[r]}
\put(43,8){$\bullet$}
\put(49,10){V}
\end{picture}}
\put(105,18){=}
\end{picture}
%<<<<<<<<<<<<<<<<<<<<<<<<<<<<<<<<<<<<<<<<<<<<<<<
\\ \nonumber
%>>>>>>>>>>>>>>>>>>>>>>>>>>>>>>>>>>>>>>>>>>>>>>>>>
\begin{picture}(105,30)(0,0)
%%%%%%%%%%%%%%%%%%%%%%%%%%%%%% box
\put(15,5){\begin{picture}(40,20)(0,0)
\put(0,15){\line(1,0){5}}
\put(0,5){\vector(1,0){10}}
\put(15,15){\vector(-1,0){10}}
\put(15,5){\line(-1,0){5}}
\put(15,0){\framebox(10,20){}}
\put(25,15){\line(1,0){5}}
\put(25,5){\vector(1,0){10}}
\put(40,15){\vector(-1,0){10}}
\put(40,5){\line(-1,0){5}}
\end{picture}}
%%%%%%%%%%%%%%%%%%%%%%%%%%%%%% half closed
\put(15,5){\begin{picture}(40,20)(0,0)
\put(40,10){\oval(10,10)[r]}
\put(43,8){$\bullet$}
\put(-13,15){$\gamma_5$}
\put(49,10){$V$}
\end{picture}}
\put(90,14){+}
\end{picture}
%<<<<<<<<<<<<<<<<<<<<<<<<<<<<<<<<<<<<<<<<<<<<<<
%>>>>>>>>>>>>>>>>>>>>>>>>>>>>>>>>>>>>>>>
\begin{picture}(75,30)(0,0)
%%%%%%%%%%%%%%%%%%%%%%%%%%%%%% box
\put(5,5){\begin{picture}(40,20)(0,0)
\put(0,15){\line(1,0){5}}
\put(0,5){\vector(1,0){10}}
\put(15,15){\vector(-1,0){10}}
\put(15,5){\line(-1,0){5}}
\put(15,0){\framebox(10,20){}}
\put(25,15){\line(1,0){5}}
\put(25,5){\vector(1,0){10}}
\put(40,15){\vector(-1,0){10}}
\put(40,5){\line(-1,0){5}}
\end{picture}}
%%%%%%%%%%%%%%%%%%%%%%%%%%%%%% half closed
\put(5,5){\begin{picture}(40,20)(0,0)
\put(40,10){\oval(10,10)[r]}
\put(43,8){$\bullet$}
\put(49,10){$\gamma_5 V$}
\end{picture}}
\end{picture}
%<<<<<<<<<<<<<<<<<<<<<<<<<<<<<<<<<<<<<<<<<<<<<<<<<<<<
\ .
\end{eqnarray}

This identity allows the reduction of terms with two axial vertices and is
the core of the present calculation.
(We remark that $S^{-1}S=I$ is introduced in (\ref{upperWI}) 
and $S$ completes the ladder in eq. (\ref{ladderDEF}) leaving the
explicit $S^{-1}$.).

%%%%%%%%%%%%%%%%%%%%%%%%%%%%%%%%%%%%%%%%%%%%%%%%%%%%%%%%%%%%%%%%
%%%%%%%%%%%%%%%%%%%%%%%%%%%%%%%%%%%%%%%%%%%%%%%%%%%%%%%%%%%%%%%%%
%%%%%%%%%%%%%%%%%%%%%%%%%%%%%%%%%%%%%%%%%%%%%%%%%%%%%%%%%%%%%%%%%
\subsection{Further Ladder Properties.}
%%%%%%%%%%%%%%%%%%%%%%%%%%%%%%%%%%%%%%%%%%%%%%%%%%%%%%%%%%%%%%%%%
We start by observing that the pseudoscalar ladder can be
Laurent-expanded around its pion pole. Keeping only the first term,
containing the pole, one obtains:
\be \label{Laddersaturation}
\begin{picture}(180,50)(0,0)
%\put(0,0){\framebox(180,50){}}
\put(40,25){\oval(10,10)[l]}
\put(20,25){$\chi_\pi$}
\put(32,22){$\bullet$}
\fermionright{40}{20}
\fermionleft{60}{30}
\put(65,22){$=$}
\put(77,22){$\frac{\chi_\pi}{c}$}
\put(92,22){$\bullet$}
\put(100,25){\oval(10,10)[l]}
\fermionleft{120}{30}
\fermionright{100}{20}
\put(120,16){\framebox(10,18){}}
\fermionleft{150}{30}
\fermionright{130}{20}
\end{picture}
\ee
where
\be
c=\frac{
\begin{picture}(75,25)(0,0)
%\put(0,0){\framebox(75,40){}}
\put(5,4){$i\chi_\pi$}
\put(22,2){$\bullet$}
\put(30,5){\oval(10,10)[l]}
\fermionleft{50}{10}
\fermionright{30}{0}
\put(50,5){\oval(10,10)[r]}
\put(60,4){$\chi_\pi$}
\put(53,2){$\bullet$}
\end{picture} } %endfraction
{p^2-m_\pi^2 } 
\ee
(in the calculations contained in this paper, $m_\pi^2=0$.)
Combining this together with the definition of $\Delta$ in
eq. (\ref{expandGamma}) one can use
\ba \label{DeltawithLadder}
\begin{picture}(180,35)(0,0)
%\put(0,0){\framebox(180,45){}}
\put(20,25){\oval(10,10)[l]}
\put(-10,25){$\frac{\Delta_j^a}{f_\pi}$}
\put(12,22){$\bullet$}
\fermionright{20}{20}
\fermionleft{40}{30}
\put(45,22){$=$}
%<<<<<<<<<<<<<<<<<<<<<<<<<<<<<<<<<<<<<<<<<<<<<<<<<<<<
%>>>>>>>>>>>>>>>>>>>>>>>>>>>>>>>>>>>>>>>>>>>>>>>>>>>>
\put(120,25){\oval(10,10)[l]}
\put(55,25){$\left( \chi_j^a -\frac{\Gamma_{A_j}^a}{i f_\pi}\right)$}
\put(112,22){$\bullet$}
\fermionright{120}{20}
\fermionleft{140}{30}
\put(145,22){$=$}
\end{picture}
\\ \nonumber
\begin{picture}(130,35)(0,0)
\put(0,22){$\left( \frac{\chi_j^a}{c_j} -\frac{\gamma_{A_j}}
{i f_\pi}\frac{\sigma^a}{2}\right)$}
\put(64,22){$\bullet$}
\put(72,25){\oval(10,10)[l]}
\fermionleft{92}{30}
\fermionright{72}{20}
\put(92,16){\framebox(10,16){}}
\fermionleft{122}{30}
\fermionright{102}{20}
\end{picture} \ .
\ea

The ladder can also be expanded in powers of the external momentum: starting
from the geometrical series (\ref{ladderDEF}) and expanding all propagators,
then resuming when possible, we can show that in analogy with the matrix relations
\ba \nonumber
\left( \frac{1}{1-y}\right)'= \left( \frac{1}{1-y}\right) y'\left( \frac{1}{1-y}\right) \\
\nonumber
\left( \frac{1}{1-y}\right)''=2\left( \frac{1}{1-y}\right)y'\left( \frac{1}{1-y}\right)
y'\left( \frac{1}{1-y}\right)+ \\ \nonumber
\left( \frac{1}{1-y}\right)y''\left( \frac{1}{1-y}\right)
\ea
one has
\ba \label{ladderEXP} \nonumber
%>>>>>>>>>>>>>>>>>>>>>>>>>>>>>>>>>>>>>>>>>>>>>>>>>>>>>>>>>>>>>>
\begin{picture}(232,30)(0,0) %(63,761)               %791-Y
%\put(0,0){\pframebox(232,30){}}
\put(0,10){                         %inicio equacao
%%%%%%%%%%%%%%%%%%%%%%%%%%%%% ladder
\begin{picture}(40,20)(0,7)
\put(0,15){\line(1,0){5}}
\put(0,5){\vector(1,0){10}}
\put(15,15){\vector(-1,0){10}}
\put(15,5){\line(-1,0){5}}
\put(15,0){\framebox(10,20){}}
\put(25,15){\line(1,0){5}}
\put(25,5){\vector(1,0){10}}
\put(40,15){\vector(-1,0){10}}
\put(40,5){\line(-1,0){5}}
\put(0,15){$k$}
\put(-5,-5){$k\! \! + \! \!P$}
\put(38,15){$q$}
\put(30,-5){$q\! \! + \! \! P$}
\end{picture}
$ \ \simeq \ $
\begin{picture}(40,20)(0,7)
\put(0,15){\line(1,0){5}}
\put(0,5){\vector(1,0){10}}
\put(15,15){\vector(-1,0){10}}
\put(15,5){\line(-1,0){5}}
\put(15,0){\framebox(10,20){}}
\put(25,15){\line(1,0){5}}
\put(25,5){\vector(1,0){10}}
\put(40,15){\vector(-1,0){10}}
\put(40,5){\line(-1,0){5}}
\put(0,15){$k$}
\put(0,5){$k$}
\put(38,15){$q$}
\put(38,5){$q$}
\end{picture}
$\ +\ $
\begin{picture}(80,20)(0,7)
\put(0,15){\line(1,0){5}}
\put(0,5){\vector(1,0){10}}
\put(15,15){\vector(-1,0){10}}
\put(15,5){\line(-1,0){5}}
\put(15,0){\framebox(10,20){}}
\put(25,15){\line(1,0){5}}
\put(25,5){\vector(1,0){10}}
\put(40,15){\vector(-1,0){10}}
\put(40,5){\line(-1,0){5}}
\put(0,15){$k$}
\put(0,5){$k$}
\put(68,15){$q$}
\put(68,5){$q$}
\put(40,0){\framebox(10,20){}}
\fermionleft{70}{15}
\fermionright{50}{5}
\put(30,15){$k'$}
\put(20,-8){$(k'\! \! +\! \! P)^{(1)}$}
\end{picture}
}
\end{picture}
%<<<<<<<<<<<<<<<<<<<<<<<<<<<<<<<<<<<<<<<<<<<<<<<<<<<<<<<
\\
+
\\  \nonumber
\begin{picture}(232,30)(0,0)
\put(0,10){
  \begin{picture}(80,20)(0,7)
  \put(0,15){\line(1,0){5}}
  \put(0,5){\vector(1,0){10}}
  \put(15,15){\vector(-1,0){10}}
  \put(15,5){\line(-1,0){5}}
  \put(15,0){\framebox(10,20){}}
  \put(25,15){\line(1,0){5}}
  \put(25,5){\vector(1,0){10}}
  \put(40,15){\vector(-1,0){10}}
  \put(40,5){\line(-1,0){5}}
  \put(0,15){$k$}
  \put(0,5){$k$}
  \put(68,15){$q$}
  \put(68,5){$q$}
  \put(40,0){\framebox(10,20){}}
  \fermionleft{70}{15}
  \fermionright{50}{5}
  \put(30,15){$k'$}
  \put(20,-8){$(k'\! \! +\! \! P)^{(2)}$}
  \end{picture}
$\ + \ $
  \begin{picture}(120,20)(0,7)
  \put(0,15){\line(1,0){5}}
  \put(0,5){\vector(1,0){10}}
  \put(15,15){\vector(-1,0){10}}
  \put(15,5){\line(-1,0){5}}
  \put(15,0){\framebox(10,20){}}
  \put(25,15){\line(1,0){5}}
  \put(25,5){\vector(1,0){10}}
  \put(50,15){\vector(-1,0){20}}
  \put(50,5){\line(-1,0){15}}
  \put(0,15){$k$}
  \put(0,5){$k$}
  \put(50,0){\framebox(10,20){}}
  \fermionleft{80}{15}
  \fermionright{60}{5}
  \put(30,15){$k'$}
  \put(15,-10){$(k'\! \! +\! \! P)^{(1)}$}
  \put(80,0){\framebox(10,20){}}
  \fermionright{90}{5}
  \fermionleft{110}{15}
  \put(111,5){$q$} \put(111,15){$q$}
  \put(55,-10){$(k''\! \! +\! \! P)^{(1)}$}
  \multiput(42,3)(0,2){6}{$\cdot$}
  \put(70,15){$k''$}
  \end{picture}
}
\end{picture}
\ea

where the following momentum expansion of the propagators is meant 
by the super indices
$^{(1)}$, $^{(2)}$
\begin{widetext}
\be \nonumber \label{prop0}
S(q+P/2)\arrowvert_{P=0} =\frac{A \slash{q}+B}{A^2 q^2 -B^2}
\ee 
\be \label{prop1}
\partial^\mu S(q+P/2)\arrowvert_{P=0} = \gamma_\mu \frac{A/2}{A^2 q^2-B^2}+
q_{\mu} \left( \frac{A' \slash{q} + B' }{A^2 q^2-B^2} -
\frac{(A \slash{q}+B)(A^2+ 2 q^2 A A' - 2 B B')}{(A^2 q^2-B^2)^2} \right)
\ee
\ba \label{prop2}
\partial^\mu \partial_\mu S(q+P/2)\arrowvert_{P=0} = \frac{\slash{q}A'}{A^2 q^2-B^2}-
\frac{A^2+2 q^2AA' -2BB'}{(A^2 q^2-B^2)^2}(\slash{q} A + 2 q^2 (\slash{q} A'+B')) \\
\nonumber
+2q^2 \left[
\frac{A'' \slash{q}+B''}{2(A^2 q^2-B^2)} + \frac{A \slash{q}+B}{(A^2 q^2-B^2)^2}
\left(
\frac{(A^2+2 q^2 A A' -2 B B')^2}{A^2 q^2-B^2}-(2 AA'+q^2(A')^2+q^2 AA'' - (B')^2-BB'')
\right) \right] \\ \nonumber
+\frac{2}{(A^2 q^2-B^2)^2}\left[ 
(A' \slash{q} + B' )(A^2 q^2-B^2)-(A \slash{q}+B)(A^2+2 q^2AA' -2BB')
\right]
\ea
\end{widetext}

The last diagram in (\ref{ladderEXP}) contains an annoying explicit rung of the
interaction. This can be eliminated at the cost of adding a diagram to any
expression where it appears: in analogy with
$$
\frac{y}{1-y}= \frac{1}{1-y}-1
$$
one can use
\begin{equation} \label{looserungkill}
%>>>>>>>>>>>>>>>>>>>>>>>>>>>>>>>>>>>>>>>>>>>>>>>>>>>>>>>>>>>>>>
\begin{picture}(232,30)(0,0) %(63,761)               %791-Y
%\put(0,0){\pframebox(232,30){}}
\put(0,10){                         %inicio equacao
%%%%%%%%%%%%%%%%%%%%%%%%%%%%% ladder
\begin{picture}(40,20)(0,7)
\put(0,15){\line(1,0){5}}
\put(0,5){\vector(1,0){10}}
\put(15,15){\vector(-1,0){10}}
\put(15,5){\line(-1,0){5}}
\put(15,0){\framebox(10,20){}}
\put(25,15){\line(1,0){5}}
\put(25,5){\vector(1,0){10}}
\put(40,15){\vector(-1,0){10}}
\put(40,5){\line(-1,0){5}}
\multiput(10,3)(0,2){6}{$\cdot$}
\end{picture}
$ \ = \ $
%%%%%%%%%%%%%%%%%%%%%%%%%% potential
\begin{picture}(40,20)(0,7)
\put(0,15){\line(1,0){5}}
\put(0,5){\vector(1,0){10}}
\put(15,15){\vector(-1,0){10}}
\put(15,5){\line(-1,0){5}}
\put(15,0){\framebox(10,20){}}
\put(25,15){\line(1,0){5}}
\put(25,5){\vector(1,0){10}}
\put(40,15){\vector(-1,0){10}}
\put(40,5){\line(-1,0){5}}
\end{picture}
$ \ - \ $
\begin{picture}(40,20)(0,7)
\put(0,15){\line(1,0){5}}
\put(0,5){\vector(1,0){10}}
\put(15,15){\vector(-1,0){10}}
\put(15,5){\line(-1,0){5}}
\end{picture}
}
\end{picture} \ .
%<<<<<<<<<<<<<<<<<<<<<<<<<<<<<<<<<<<<<<<<<<<<<<<<<<<<<<<
\end{equation}

%%%%%%%%%%%%%%%%%%%%%%%%%%%%%%%%%%%%%%%%%%%%%%%%%%%%%%%%%%%%%%%%%
%%%%%%%%%%%%%%%%%%%%%%%%%%%%%%%%%%%%%%%%%%%%%%%%%%%%%%%%%%%%%%%%%%
%%%%%%%%%%%%%%%%%%%%%%%%%%%%%%%%%%%%%%%%%%%%%%%%%%%%%%%%%%%%%%%%%
%%%%%%%%%%%%%%%%%%%%%%%%%%%%%%%%%%%%%%%%%%%%%%%%%%%%%%%%%%%%%%%%%%
\section{Pion-Pion scattering} \label{pionpion}
%%%%%%%%%%%%%%%%%%%%%%%%%%%%%%%%%%%%%%%%%%%%%%%%%%%%%%%%%%%%%%%%%%
\subsection{Generalities.}
%%%%%%%%%%%%%%%%%%%%%%%%%%%%%%%%%%%%%%%%%%%%%%%%%%%%%%%%%%%%%%%%%%

The pion scattering amplitude with the Bethe-Salpeter and planar
 approximations can be derived from
the following Feynman diagrams:

\begin{equation} \label{PIPISCATT}
%>>>>>>>>>>>>>>>>>>>>>>>>>>>>>>>>>>>>>>>>>>>>>>
\begin{picture}(45,65)(0,0)
%%%%%%%%%%%%%%%%%%%%%%%%%%%%%% box with two gammas
\put(10,13){\begin{picture}(20,40)(0,0)
\put(5,0){\line(0,1){5}}
\put(5,15){\vector(0,-1){10}}
\put(15,15){\line(0,-1){5}}
\put(15,0){\vector(0,1){10}}
\put(0,15){\framebox(20,10){}}
\put(5,25){\line(0,1){5}}
\put(5,40){\vector(0,-1){10}}
\put(15,40){\line(0,-1){5}}
\put(15,25){\vector(0,1){10}}
\end{picture}}
\put(10,8){\begin{picture}(20,40)(0,0)
\put(10,40){\oval(10,26)[t]}
\put(2,42){$\bullet$}
\put(-11,48){$\chi_{\pi_4}$}
\put(13,42){$\bullet$}
\put(18,48){$\chi_{\pi_1}$}
\put(10,53){\vector(-1,0){2}}
\end{picture}}
\put(10,-27){\begin{picture}(20,40)(0,0)
\put(10,45){\oval(10,26)[b]}
\put(2,38){$\bullet$}
\put(-11,34){$\chi_{\pi_3}$}
\put(13,38){$\bullet$}
\put(18,34){$\chi_{\pi_2}$}
\put(10,32){\vector(1,0){2}}
\end{picture}}
\put(40,30){+}
\end{picture}
%<<<<<<<<<<<<<<<<<<<<<<<<<<<<<<<<<<<<<<<<<<<<<<<<<<<
%>>>>>>>>>>>>>>>>>>>>>>>>>>>>>>>>>>>>>>>>>>>>>>
\begin{picture}(90,50)(0,0)
\put(0,11){
\begin{picture}(90,40)(0,0)
%%%%%%%%%%%%%%%%%%%%%%%%%%%%%% box with two gammas
\put(25,10){\begin{picture}(40,20)(0,0)
\put(0,15){\line(1,0){5}}
\put(0,5){\vector(1,0){10}}
\put(15,15){\vector(-1,0){10}}
\put(15,5){\line(-1,0){5}}
\put(15,0){\framebox(10,20){}}
\put(25,15){\line(1,0){5}}
\put(25,5){\vector(1,0){10}}
\put(40,15){\vector(-1,0){10}}
\put(40,5){\line(-1,0){5}}
\end{picture}}
\put(25,10){\begin{picture}(40,20)(0,0)
\put(40,10){\oval(30,10)[r]}
\put(42,2){$\bullet$}
\put(48,-4){$\chi_{\pi_2}$}
\put(42,13){$\bullet$}
\put(48,20){$\chi_{\pi_1}$}
\put(55,10){\vector(0,1){2}}
\end{picture}}
\put(-20,10){\begin{picture}(40,20)(0,0)
\put(45,10){\oval(30,10)[l]}
\put(38,2){$\bullet$}
\put(34,-4){$\chi_{\pi_3}$}
\put(38,13){$\bullet$}
\put(34,20){$\chi_{\pi_4}$}
\put(30,10){\vector(0,-1){2}}
\end{picture}}
\end{picture}
}
\end{picture}
%<<<<<<<<<<<<<<<<<<<<<<<<<<<<<<<<<<<<<<<<<<<<<<<<<<<
%>>>>>>>>>>>>>>>>>>>>>>>>>>>>>>>>>>>>>>>>>>>>>>>>>>>
\begin{picture}(65,50)(0,0)
\put(0,11){
\begin{picture}(65,40)(0,0)
%%%%%%%%%%%%%%%%%%%%%%%%%%%%%% four Gammas
\put(0,15){$-$}
\put(25,10){\begin{picture}(40,20)(0,0)
\put(2,15){\line(1,0){5}}
\put(0,5){\vector(1,0){10}}
\put(15,15){\vector(-1,0){10}}
\put(15,5){\line(-1,0){5}}
\end{picture}}
\put(0,10){\begin{picture}(40,20)(0,0)
\put(40,10){\oval(30,10)[r]}
\put(42,2){$\bullet$}
\put(48,-4){$\chi_{\pi_2}$}
\put(42,13){$\bullet$}
\put(48,20){$\chi_{\pi_1}$}
\put(55,10){\vector(0,1){2}}
\end{picture}}
\put(-20,10){\begin{picture}(40,20)(0,0)
\put(45,10){\oval(30,10)[l]}
\put(38,2){$\bullet$}
\put(34,-4){$\chi_{\pi_3}$}
\put(38,13){$\bullet$}
\put(34,20){$\chi_{\pi_4}$}
\put(30,10){\vector(0,-1){2}}
\end{picture}}
\end{picture} }  %end put
\end{picture}
%<<<<<<<<<<<<<<<<<<<<<<<<<<<<<<<<<<<<<<<<<<<<<<<<<<<
\end{equation}
where the first two terms provide all possible planar topologies, but upon
substitution of (\ref{ladderDEF}) their zeroth order is seen to be double
counted, hence we subtract it. The two approximations involved are: first,
coupling of the pion to higher Fock space states is not considered and second, 
only planar diagrams are utilized. 
It was shown, using the axial Ward Identity, 
that these two approximations are consistent within the Schwinger-Dyson 
method in the rainbow approximation for the fermion mass generation
and in the ladder approximation for the bound state equation
\cite{weall,Pedrosolo}. 
This is also consistent with past work \cite{Pich} on resonance exchange, and
is equivalent to  the lowest order in a $1/N_c$ expansion.
Reduction of this combination of Feynman diagrams to $O(p^4)$, $O({M_\pi}^0)$ is
our goal.
The calculations in this section will treat the $P$'s as incoming momenta.
Matching the dummy $P_j$ to the incoming $q_{i_1}$, $q_{i_2}$ and outgoing
$q_{o_1}$, $q_{o_2}$ physical pion momenta leads to 6 different
permutations, namely
\be \label{permutations}
(P_1,P_2,P_3,P_4)=
\left(  \begin{tabular}{c}
$(q_{i1},\ q_{i2}, -q_{o2}, -q_{o1})$ \\
$(q_{i1},\ q_{i2}, -q_{o1}, -q_{o2})$ \\
$(q_{i1}, -q_{o1}, -q_{o2},\ q_{i2})$ \\
$(q_{i1}, -q_{o2}, -q_{o1},\ q_{i2})$ \\
$(q_{i1}, -q_{o1},\ q_{i2}, -q_{o2})$ \\
$(q_{i1}, -q_{o2},\ q_{i2}, -q_{o1})$ \\
\end{tabular} \right) \ ,
\ee
where the first momentum is fixed to avoid double counting by rotational
symmetry of the $\pi$-$\pi$ scattering amplitude (\ref{PIPISCATT}).

We concentrate on $A(s,t,u)$, the amplitude for 
$\pi_+ \pi_- \longrightarrow \pi_0 \pi_0$, and use the following isospin
wavefunctions
\ba \label{BSisospin}
\pi_+= \frac{1}{2}(\sigma_1 + i \sigma_2)\\ \nonumber
\pi_-= \frac{1}{2}(-\sigma_1 + i \sigma_2)\\ \nonumber
\pi_0 = \frac{1}{\sqrt{2}} \sigma_3 \ .
\ea
From them follow the traces
\be \label{isospinfirst}
Tr(\pi_+ \pi_- \pi_0 \pi_0) = -\frac{1}{2}
\ee
(which multiplies the four first permutations in (\ref{permutations}) above
where the two charged pions are adjacent) and
\be \label{isospinsecond}
Tr(\pi_+ \pi_0 \pi_- \pi_0)= \frac{1}{2}
\ee
(multiplying the last two permutations in (\ref{permutations}) where the two
charged pions are in opposite corners of the amplitude).

Finally the Mandelstam variables in the chiral limit satisfy
\ba
s= 2q_{i1} q_{i2}=  2q_{o1} q_{o2} \\ \nonumber
t=-2q_{i1} q_{o1}= -2q_{i2} q_{o2} \\ \nonumber
u=-2q_{i1} q_{o2}= -2q_{i2} q_{o1} \ .
\ea

Since isospin has been factored out, we can ignore it in the rest
of the calculation. Instead of working with $\Gamma_a$ we employ
$\Gamma$ as defined in eq. (\ref{GAvertex2}). Accordingly, we define
\be
\chi^a := \frac{\sigma^a}{\sqrt{2}} \chi
\ee
to yield the normalized isospin wavefunctions in (\ref{BSisospin})
and the normalization for $\chi$, $\Gamma_A$ will be
\be \label{expandGammabis}
\chi=\frac{-i \Gamma_A + \Delta}{\sqrt{2} f_\pi} \ .
\ee

We now start evaluating the Feynman amplitude in terms of the
$P$'s.
Start by employing (\ref{expandGammabis}) to treat the product of four BS
amplitudes:

\ba
\chi_{\pi_1} \chi_{\pi_2} \chi_{\pi_3} \chi_{\pi_4} = \, \, \, \, \, \,
\, \, \, \, \\ \nonumber \\ \nonumber 
\left( \frac{1}{\sqrt{2} f_\pi}
\right)^4 \! \! \!
\left( \frac{\Gamma_1}{i}+ \Delta_1 \right) \! \! \!
\left( \frac{\Gamma_2}{i}+\Delta_2 \right) \! \! \!
\left( \frac{\Gamma_3}{i}+\Delta_3 \right) \! \! \!
\left( \frac{\Gamma_4}{i}+\Delta_4 \right) \! \! \!
 \\ \nonumber 
=
\left(  \frac{1}{\sqrt{2} f_\pi} \right)^4 \! \! 
\left(  \Delta_1 \Delta_2 \Delta_3 \Delta_4
+ \frac{\Gamma_1}{i} \Delta_2 \Delta_3
\Delta_4 + {\rm perm.} \right) + ...
\ea

where the omitted terms, contain an increasing number of powers of
$\Gamma$. Next we proceed to a term by term analysis of this 
expansion, further explained in the paper \cite{Pedrosolo}.

%%%%%%%%%%%%%%%%%%%%%%%%%%%%%%%%%%%%%%%%%%%%%%%%%%%%%%%%%%%%%%%%%%%%%%
%%%%%%%%%%%%%%%%%%%%%%%%%%%%%%%%%%%%%%%%%%%%%%%%%%%%%%%%%%%%%%%%%%%%%%
%%%%%%%%%%%%%%%%%%%%%%%%%%%%%%%%%%%%%%%%%%%%%%%%%%%%%%%%%%%%%%%%%%%%%%%
\subsection{Contribution with 4-$\Delta$s.}
%%%%%%%%%%%%%%%%%%%%%%%%%%%%%%%%%%%%%%%%%%%%%%%%%%%%%%%%%%%%%%%%%%%%%%
The term with four $\Delta$'s is model dependent and no chiral properties
can be used to simplify it, since it contains corrections to the BS
pion wavefunction beyond the zero momentum limit. Without evaluating it
explicitly in a particular model yet (but see later), we can parameterize it. To the order
$m^0$ we work here, for on-shell pions, $P_i^2=0$ for all $i$. Therefore,
the 4-$\Delta$ diagrams can only be a combination of products of
different momenta, $P_i P_j$. But to order $P^4$, since each $\Delta$
brings at least one momentum power (the zeroth power is accounted for
already in
$\Gamma$), only combinations of the type $(P_1P_2)(P_3P_4)$,
$(P_1P_4)(P_2P_3)$, $(P_1P_3)(P_2P_4)$ can appear. The coefficients of the
first two
terms have to be equal because of the cyclic symmetry of eqn.
(\ref{PIPISCATT}) The coefficient of the last term is in general
independent.

In terms of fictitious momenta, all flowing into the diagram, whose
conservation law is $P_1+P_2+P_3+P_4=0$, we obtain

\ba \label{defined1d2}
%>>>>>>>>>>>>>>>>>>>>>>>>>>>>>>>>>>>>>>>>>>>>>>
\begin{picture}(50,65)(0,0)
%%%%%%%%%%%%%%%%%%%%%%%%%%%%%% box with two gammas
\put(10,13){\begin{picture}(20,40)(0,0)
\put(5,0){\line(0,1){5}}
\put(5,15){\vector(0,-1){10}}
\put(15,15){\line(0,-1){5}}
\put(15,0){\vector(0,1){10}}
\put(0,15){\framebox(20,10){}}
\put(5,25){\line(0,1){5}}
\put(5,40){\vector(0,-1){10}}
\put(15,40){\line(0,-1){5}}
\put(15,25){\vector(0,1){10}}
\end{picture}}
\put(10,8){\begin{picture}(20,40)(0,0)
\put(10,40){\oval(10,26)[t]}
\put(2,42){$\bullet$}
\put(-9,48){$\Delta_4$}
\put(13,42){$\bullet$}
\put(18,48){$\Delta_1$}
\put(10,53){\vector(-1,0){2}}
\end{picture}}
\put(10,-27){\begin{picture}(20,40)(0,0)
\put(10,45){\oval(10,26)[b]}
\put(2,38){$\bullet$}
\put(-9,34){$\Delta_3$}
\put(13,38){$\bullet$}
\put(18,34){$\Delta_2$}
\put(10,32){\vector(1,0){2}}
\end{picture}}
\put(40,30){+}
\end{picture}
%<<<<<<<<<<<<<<<<<<<<<<<<<<<<<<<<<<<<<<<<<<<<<<<<<<<
%>>>>>>>>>>>>>>>>>>>>>>>>>>>>>>>>>>>>>>>>>>>>>>
\begin{picture}(100,50)(0,0) \put(0,11){
\begin{picture}(100,40)(0,0)
%%%%%%%%%%%%%%%%%%%%%%%%%%%%%% box with two gammas
\put(25,10){\begin{picture}(40,20)(0,0)
\put(0,15){\line(1,0){5}}
\put(0,5){\vector(1,0){10}}
\put(15,15){\vector(-1,0){10}}
\put(15,5){\line(-1,0){5}}
\put(15,0){\framebox(10,20){}}
\put(25,15){\line(1,0){5}}
\put(25,5){\vector(1,0){10}}
\put(40,15){\vector(-1,0){10}}
\put(40,5){\line(-1,0){5}}
\end{picture}}
\put(25,10){\begin{picture}(40,20)(0,0)
\put(40,10){\oval(30,10)[r]}
\put(42,2){$\bullet$}
\put(48,-4){$\Delta_2$}
\put(42,13){$\bullet$}
\put(48,20){$\Delta_1$}
\put(55,10){\vector(0,1){2}}
\end{picture}}
\put(-20,10){\begin{picture}(40,20)(0,0)
\put(45,10){\oval(30,10)[l]}
\put(38,2){$\bullet$}
\put(34,-4){$\Delta_3$}
\put(38,13){$\bullet$}
\put(34,20){$\Delta_4$}
\put(30,10){\vector(0,-1){2}}
\end{picture}}
\end{picture} } %end put
\put(95,30){$-$}
\end{picture}
%<<<<<<<<<<<<<<<<<<<<<<<<<<<<<<<<<<<<<<<<<<<<<<<<<<<
%>>>>>>>>>>>>>>>>>>>>>>>>>>>>>>>>>>>>>>>>>>>>>>>>>>>
\begin{picture}(65,50)(0,0) \put(0,11){
\begin{picture}(65,40)(0,0)
%%%%%%%%%%%%%%%%%%%%%%%%%%%%%% four Gammas
\put(25,10){\begin{picture}(40,20)(0,0)
\put(0,15){\line(1,0){5}}
\put(0,5){\vector(1,0){10}}
\put(15,15){\vector(-1,0){10}}
\put(15,5){\line(-1,0){5}}
\end{picture}}
\put(0,10){\begin{picture}(40,20)(0,0)
\put(40,10){\oval(30,10)[r]}
\put(42,2){$\bullet$}
\put(48,-4){$\Delta_2$}
\put(42,13){$\bullet$}
\put(48,20){$\Delta_1$}
\put(55,10){\vector(0,1){2}}
\end{picture}}
\put(-20,10){\begin{picture}(40,20)(0,0)
\put(45,10){\oval(30,10)[l]}
\put(38,2){$\bullet$}
\put(34,-4){$\Delta_3$}
\put(38,13){$\bullet$}
\put(34,20){$\Delta_4$}
\put(30,10){\vector(0,-1){2}}
\end{picture}}
\end{picture} } %end put
\end{picture}
%<<<<<<<<<<<<<<<<<<<<<<<<<<<<<<<<<<<<<<<<<<<<<<<<<<<
\\ \nonumber
= 3 d_1 (P_1\cd P_2 \ P_3 \cd P_4+ P_1 \cd P_4 \ P_2 \cd P_3) + 3 d_2 P_1 \cd P_3 \ P_2 \cd P_4  \ .
\ea
The two numbers $d_1$, $d_2$ contain the non-trivial information in this
diagram. We have explicitly pulled out the color factor (3 as will be shown shortly)
from the $d's$, which contain in this way only momentum and spin (since flavor will
be dealt with at the end when the external legs are matched to the physical particles).

The third term in (\ref{defined1d2}), to order 4 in momentum, with no ladder, 
is simply a wavefunction overlap given by the usual Feynman rules 
(notice an extra (-1) will be due to the fermion loop)
\begin{widetext}
\ba
-\int \frac{d^4q}{(2\pi)^4} \left(
\frac{i}{A^2 q^2-B^2} \right)^4
Tr \left[
\Delta^{(1)}(P_1,q) (A \slash{q}+B) \Delta^{(1)}(P_4,q)
 (A \slash{q}+B)\Delta^{(1)}(P_3,q) (A \slash{q}+B)
\Delta^{(1)}(P_2,q)(A \slash{q}+B) \right] \ .
\ea
\end{widetext}
The Dirac traces can easily be computed with FORM. The integral
is then reduced by using tensor identities like
$$
\int F(q^2) q^\mu q^\nu q^\rho q^\sigma = \frac{g^{\mu \nu}
g^{\rho \sigma} + g^{\mu \rho} g^{\nu \sigma} + g^{\mu \sigma}
g^{\nu \rho}}{24} \int q^4 F(q^2)
$$
to a one dimensional expression which can then be numerically evaluated
once a specific model (and hence Bethe-Salpeter wavefunctions) is
chosen. Notice that this simple momentum routing is correct only to order
$P^4$.

The first and second diagram in (\ref{defined1d2}) contain a ladder. 
A simple way to calculate them is to write an integral equation for the object

\ba \label{Definesqcup}
i \sqcup :=
%>>>>>>>>>>>>>>>>>>>>>>>>>>>>>>>>>>>>>>>>>>>>>>
\begin{picture}(100,50)(-25,25) \put(0,11){
\begin{picture}(100,40)(0,0)
%%%%%%%%%%%%%%%%%%%%%%%%%%%%%% box with two gammas
\put(25,10){\begin{picture}(40,20)(0,0)
\put(10,15){\line(1,0){5}}
%\put(0,5){\vector(1,0){10}}
\put(6,3){x}
\put(6,13){x}
%\put(15,15){\vector(-1,0){10}}
\put(15,5){\line(-1,0){5}}
\put(15,0){\framebox(10,20){}}
\put(25,15){\line(1,0){5}}
\put(25,5){\vector(1,0){10}}
\put(40,15){\vector(-1,0){10}}
\put(40,5){\line(-1,0){5}}
\end{picture}}
\put(25,10){\begin{picture}(40,20)(0,0)
\put(40,10){\oval(30,10)[r]}
\put(42,2){$\bullet$}
\put(48,-4){$\Delta_2$}
\put(42,13){$\bullet$}
\put(48,20){$\Delta_1$}
\put(55,10){\vector(0,1){2}}
\end{picture}}
\end{picture} } %end put
\end{picture}
\ea
whose most general expansion up to second order in the external
pion momenta is
\ba \label{expandsqcup}
\sqcup := \sqcup_0(k^2) P_1 \cd P_2           +
          \sqcup_1(k^2) k \cd P_1 \ k \cd P_2 + \\ \nonumber
          \sqcup_2(k^2) P_1 \cd P_2 \slash{k} +
          \sqcup_3(k^2) k \cd P_1 \ k \cd P_2 \slash{k} +
          \sqcup_4(k^2) k \cd P_2 \slash{P}_1 +  \\ \nonumber
          \sqcup_5(k^2) k \cd P_1 \slash{P}_2 + 
          \sqcup_6(k^2) k \cd P_2 \slash{k} \slash{P}_1 +
          \sqcup_7(k^2) k \cd P_1 \slash{k}\slash{P}_2 + \\ \nonumber
          \sqcup_8(k^2) \slash{P}_1 \slash{P}_2 +
          \sqcup_9(k^2) \slash{k} \slash{P}_1 \slash{P}_2 \ .
\ea
The functions $\sqcup_i$ are obtained by projecting this linear integral
equation (analogous to the Bethe-Salpeter equation)

\ba \label{eqnforsqcup}
i \sqcup(P1,P2,k) =
%>>>>>>>>>>>>>>>>>>>>>>>>>>>>>>>>>>>>>>>>>>>>>>
\begin{picture}(100,50)(25,25) \put(0,11){
\begin{picture}(50,40)(0,0)
%%%%%%%%%%%%%%%%%%%%%%%%%%%%%% box with two gammas
\put(25,10){\begin{picture}(40,20)(0,0)
\put(10,15){\line(1,0){5}}
%\put(0,5){\vector(1,0){10}}
\put(6,3){x}
\put(6,13){x}
%\put(15,15){\vector(-1,0){10}}
\put(15,5){\line(-1,0){5}}
\end{picture}}
\put(10,10){\begin{picture}(40,20)(0,0)
\put(30,10){\oval(30,10)[r]}
\put(32,2){$\bullet$}
\put(38,-4){$\Delta_2$}
\put(32,13){$\bullet$}
\put(38,20){$\Delta_1$}
\put(45,10){\vector(0,1){2}}
\end{picture}}
\end{picture} } %end put
\put(70,30){+}
\put(60,11){
\begin{picture}(50,40)(0,0)
%%%%%%%%%%%%%%%%%%%%%%%%%%%%%% box with two gammas
\put(25,10){\begin{picture}(40,20)(0,0)
\put(10,15){\line(1,0){5}}
%\put(0,5){\vector(1,0){10}}
\put(6,3){x}
\put(6,13){x}
%\put(15,15){\vector(-1,0){10}}
\put(15,5){\line(-1,0){5}}
\end{picture}}
\put(10,10){\begin{picture}(40,20)(0,0)
\put(30,10){\oval(30,10)[r]}
\put(42,7){*}
\put(50,8){$i \sqcup$}
\multiput(30,3)(0,2){6}{$\cdot$}
\end{picture}}
\end{picture} } %end put

\end{picture}
\ea
with the matrix projectors $I$, $\slash{k}$,$\slash{k}\slash{P}_1$, ...
$\slash{k} \slash{P}_1 \slash{P}_2$, which provides us with a
linear system of eight integral equations for the $\sqcup_i$.
Defining a convenient quantity $D:=2 k \cd P_1 \ k\cd P_2 -k^2 P_1 \cd P_2$
the projections are
\begin{widetext}
\ba \label{Linearsqcup}
4 D (P_1\cd P_2 \sqcup_0 + k\cd P_1 \ k\cd P_2 \sqcup_1)&=&Tr[(2D - 2 k\cd P_1 \slash{k}
\slash{P_2} + q^2 \slash{P}_1 \slash{P}_2)\sqcup] \\ \nonumber
4 D (P_1\cd P_2 \sqcup_2 + k\cd P_1 \ k\cd P_2 \sqcup_3)&=&Tr[(-2 P_1\cd P_2 \slash{k}
+2 k\cd P_2 \slash{P}_1 +\slash{k}\slash{P}_1 \slash{P}_2)\sqcup] \\ \nonumber
4 P_1 \cd P_2 D k\cd P_2 \sqcup_4 &=& Tr[(2 k\cd P_2 \ P_1\cd P_2 \slash{k} -2 (k\cd P_2)^2 \slash{P}_1
+D \slash{P}_2 - k\cd P_2  \slash{k}\slash{P}_1 \slash{P}_2)\sqcup] \\ \nonumber
4 P_1 \cd P_2 D k\cd P_1 \sqcup_5 &=& Tr[(D \slash{P}_1-2 (k\cd P_1)^2+k\cd P_1\slash{k}\slash{P}_1
 \slash{P}_2 )\sqcup] \\ \nonumber
4 P_1\cd P_2 D k\cd P_2 \sqcup_6 &=&Tr[(P_1 \cd P_2 \slash{k}  \slash{P}_2 -k\cd P_2
\slash{P}_1 \slash{P}_2 )\sqcup] \\ \nonumber
4 P_1\cd P_2 D k\cd P_1 \sqcup_7 &=& Tr[(-2 k\cd P_1 \ P_1\cd P_2 + P_1 \cd P_2  \slash{k}\slash{P}_1
+k\cd P_1\slash{k}\slash{P}_1 \slash{P}_2 )\sqcup] \\ \nonumber
4 P_1 \cd P_2 D \sqcup_8 &=& Tr[(k^2 P_1 \cd P_2 - k\cd P_2 \slash{k}\slash{P}_1 +
k\cd P_1 \slash{k} \slash{P}_2 - k^2 \slash{P}_1 \slash{P}_2)\sqcup] \\ \nonumber
4 P_1 \cd P_2 D \sqcup_9 &=& Tr[(P_1\cd P_2 \slash{k}-k\cd P_2 \slash{P}_1 +k\cd P_1 \slash{P}_2
-\slash{k}\slash{P}_1 \slash{P}_2)\sqcup]
\ea
\end{widetext}
and the inhomogeneous part of the equation can easily be written down from the first RHS diagram in
(\ref{eqnforsqcup}). Once the equation for $\sqcup$ is solved in a computer, the diagram can be closed
from the left to give
\be \label{closedelta4}
\begin{picture}(60,60)(0,0)
\put(-20,10){\begin{picture}(40,20)(0,0)
\put(45,10){\oval(30,10)[l]}
\put(38,2){$\bullet$}
\put(34,-4){$\Delta_3$}
\put(38,13){$\bullet$}
\put(34,20){$\Delta_4$}
\put(30,10){\vector(0,-1){2}}
\put(40,10){\oval(30,10)[r]}
\put(53,9){$\bullet$}
\put(57,8){$i \sqcup_{1,2}$}
\end{picture}}
\end{picture}
\ee
and calculated as a simple integral. Here the color factor of 3 in 
(\ref{defined1d2}) can be easily seen, since all three vertices
are color singlets and carry $\delta_{cc'}$ in color space..

%%%%%%%%%%%%%%%%%%%%%%%%%%%%%%%%%%%%%%%%%%%%%%%%%%%%%%%%%%%%%%%%+
%%%%%%%%%%%%%%%%%%%%%%%%%%%%%%%%%%%%%%%%%%%%%%%%%%%%%%%%%%%%%%%%
%%%%%%%%%%%%%%%%%%%%%%%%%%%%%%%%%%%%%%%%%%%%%%%%%%%%%%%%%%%%%%%%
\subsection{Contribution with $\Gamma - 3\Delta$.}
%%%%%%%%%%%%%%%%%%%%%%%%%%%%%%%%%%%%%%%%%%%%%%%%%%%%%%%%%%%%%%%

We will reduce the $3-\Delta$ contribution fixing the indices $\Gamma_4
\Delta_1 \Delta_2 \Delta_3$ (the other permutations can at the end be
easily generated). Employing (\ref{DeltawithLadder})
in

\be \label{OneGThreeD}
%>>>>>>>>>>>>>>>>>>>>>>>>>>>>>>>>>>>>>>>>>>>>>>
\begin{picture}(50,65)(0,0)
%%%%%%%%%%%%%%%%%%%%%%%%%%%%%% box with two gammas
\put(10,13){\begin{picture}(20,40)(0,0)
\put(5,0){\line(0,1){5}}
\put(5,15){\vector(0,-1){10}}
\put(15,15){\line(0,-1){5}}
\put(15,0){\vector(0,1){10}}
\put(0,15){\framebox(20,10){}}
\put(5,25){\line(0,1){5}}
\put(5,40){\vector(0,-1){10}}
\put(15,40){\line(0,-1){5}}
\put(15,25){\vector(0,1){10}}
\end{picture}}
\put(10,8){\begin{picture}(20,40)(0,0)
\put(10,40){\oval(10,26)[t]}
\put(2,42){$\bullet$}
\put(-9,48){$\Gamma_4$}
\put(13,42){$\bullet$}
\put(18,48){$\Delta_1$}
\put(10,53){\vector(-1,0){2}}
\end{picture}}
\put(10,-27){\begin{picture}(20,40)(0,0)
\put(10,45){\oval(10,26)[b]}
\put(2,38){$\bullet$}
\put(-9,34){$\Delta_3$}
\put(13,38){$\bullet$}
\put(18,34){$\Delta_2$}
\put(10,32){\vector(1,0){2}}
\end{picture}}
\put(40,30){+}
\end{picture}
%<<<<<<<<<<<<<<<<<<<<<<<<<<<<<<<<<<<<<<<<<<<<<<<<<<<
%>>>>>>>>>>>>>>>>>>>>>>>>>>>>>>>>>>>>>>>>>>>>>>
\begin{picture}(100,50)(0,0) \put(0,11){
\begin{picture}(100,40)(0,0)
%%%%%%%%%%%%%%%%%%%%%%%%%%%%%% box with two gammas
\put(25,10){\begin{picture}(40,20)(0,0)
\put(0,15){\line(1,0){5}}
\put(0,5){\vector(1,0){10}}
\put(15,15){\vector(-1,0){10}}
\put(15,5){\line(-1,0){5}}
\put(15,0){\framebox(10,20){}}
\put(25,15){\line(1,0){5}}
\put(25,5){\vector(1,0){10}}
\put(40,15){\vector(-1,0){10}}
\put(40,5){\line(-1,0){5}}
\end{picture}}
\put(25,10){\begin{picture}(40,20)(0,0)
\put(40,10){\oval(30,10)[r]}
\put(42,2){$\bullet$}
\put(48,-4){$\Delta_2$}
\put(42,13){$\bullet$}
\put(48,20){$\Delta_1$}
\put(55,10){\vector(0,1){2}}
\end{picture}}
\put(-20,10){\begin{picture}(40,20)(0,0)
\put(45,10){\oval(30,10)[l]}
\put(38,2){$\bullet$}
\put(34,-4){$\Delta_3$}
\put(38,13){$\bullet$}
\put(34,20){$\Gamma_4$}
\put(30,10){\vector(0,-1){2}}
\end{picture}}
\end{picture} } %end put
\put(95,30){$-$}
\end{picture}
%<<<<<<<<<<<<<<<<<<<<<<<<<<<<<<<<<<<<<<<<<<<<<<<<<<<
%>>>>>>>>>>>>>>>>>>>>>>>>>>>>>>>>>>>>>>>>>>>>>>>>>>>
\begin{picture}(65,50)(0,0) \put(0,11){
\begin{picture}(65,40)(0,0)
%%%%%%%%%%%%%%%%%%%%%%%%%%%%%% four Gammas
\put(25,10){\begin{picture}(40,20)(0,0)
\put(0,15){\line(1,0){5}}
\put(0,5){\vector(1,0){10}}
\put(15,15){\vector(-1,0){10}}
\put(15,5){\line(-1,0){5}}
\end{picture}}
\put(0,10){\begin{picture}(40,20)(0,0)
\put(40,10){\oval(30,10)[r]}
\put(42,2){$\bullet$}
\put(48,-4){$\Delta_2$}
\put(42,13){$\bullet$}
\put(48,20){$\Delta_1$}
\put(55,10){\vector(0,1){2}}
\end{picture}}
\put(-20,10){\begin{picture}(40,20)(0,0)
\put(45,10){\oval(30,10)[l]}
\put(38,2){$\bullet$}
\put(34,-4){$\Delta_3$}
\put(38,13){$\bullet$}
\put(34,20){$\Gamma_4$}
\put(30,10){\vector(0,-1){2}}
\end{picture}}
\end{picture} } %end put
\end{picture}
%<<<<<<<<<<<<<<<<<<<<<<<<<<<<<<<<<<<<<<<<<<<<<<<<<<<
\ee
(in the first diagram fix $j=3$ to substitute (\ref{DeltawithLadder}), in
the second diagram employ $j=1$) to obtain

\ba \label{OneGThreeD2}
\begin{picture}(200,45)(0,0)
\put(0,22){$\left(\! \frac{\sqrt{2} f_\pi \chi_3}{c_3}
\! -\! \frac{\gamma_{A_3}}{i}\!  \right)$}
\put(70,25){\oval(10,10)[l]} \put(62,23){$\bullet$}
\fermionleft{90}{30} \fermionright{70}{20}
\put(90,16){\framebox(10,18){}}
\fermionleft{120}{30} \fermionright{100}{20}
\put(113,18){$\bullet$} \put(109,8){$S^{-1}$}
\put(113,27){$\bullet$} \put(109,35){$\Gamma_4$}
\fermionleft{130}{30} \fermionright{110}{20}
\put(130,16){\framebox(10,18){}}
\fermionleft{160}{30} \fermionright{140}{20}
\put(160,25){\oval(10,10)[r]}
\put(160,27){$\bullet$} \put(165,35){$\Delta_1$}
\put(160,18){$\bullet$} \put(165,8){$\Delta_2$}
\put(170,22){$+$}
\end{picture}
%>>>>>>>>>>>>>>>>>>>>>>>>>>>>>>>>>>>>>>>>>>>>>
\\ \nonumber
%<<<<<<<<<<<<<<<<<<<<<<<<<<<<<<<<<<<<<<<<<<<<<
\begin{picture}(200,45)(0,0)
\put(0,22){$\left(\! \frac{\sqrt{2} f_\pi \chi_1}{c_1}
\! -\! \frac{\gamma_{A_3}}{i}\!  \right)$}
\put(70,25){\oval(10,10)[l]} \put(62,23){$\bullet$}
\fermionleft{90}{30} \fermionright{70}{20}
\put(90,16){\framebox(10,18){}}
\fermionleft{120}{30} \fermionright{100}{20}
\put(113,18){$\bullet$} \put(109,8){$\Gamma_4$}
\put(113,27){$\bullet$} \put(109,35){$S^{-1}$}
\fermionleft{130}{30} \fermionright{110}{20}
\put(130,16){\framebox(10,18){}}
\fermionleft{160}{30} \fermionright{140}{20}
\put(160,25){\oval(10,10)[r]}
\put(160,27){$\bullet$} \put(165,35){$\Delta_2$}
\put(160,18){$\bullet$} \put(165,8){$\Delta_3$}
\put(170,22){$-$}
\end{picture}
%>>>>>>>>>>>>>>>>>>>>>>>>>>>>>>>>>>>>>>>>>>>>>
\\ \nonumber
%<<<<<<<<<<<<<<<<<<<<<<<<<<<<<<<<<<<<<<<<<<<<<
\begin{picture}(200,45)(0,0)
\put(30,25){\oval(10,10)[l]}
\put(30,27){$\bullet$} \put(22,35){$\Gamma_4$}
\put(30,18){$\bullet$} \put(22,8){$\Delta_3$}
\fermionleft{50}{30} \fermionright{30}{20}
\put(50,25){\oval(10,10)[r]}
\put(50,27){$\bullet$} \put(55,35){$\Delta_1$}
\put(50,18){$\bullet$} \put(55,8){$\Delta_2$}
\end{picture}
\ea

Next one can apply (\ref{upperWI2}) to the single $\Gamma$ appearing in this
expression to generate, after some simple manipulations:
\ba
\begin{picture}(200,45)(0,0)
\put(0,22){$\gamma_5 \left(\! \frac{\sqrt{2} f_\pi \chi_3}{c_3}
\! -\! \frac{\gamma_{A_3}}{i}\!  \right)$}
\put(80,25){\oval(10,10)[l]} \put(72,23){$\bullet$}
\fermionleft{100}{30} \fermionright{80}{20}
\put(100,16){\framebox(10,18){}}
\fermionleft{130}{30} \fermionright{110}{20}
\put(130,25){\oval(10,10)[r]}
\put(130,27){$\bullet$} \put(135,35){$\Delta_1$}
\put(130,18){$\bullet$} \put(135,8){$\Delta_2$}
\put(160,22){$+$}
\end{picture}
%>>>>>>>>>>>>>>>>>>>>>>>>>>>>>>>>>>>>>>>>>>>>>>>>>>>>>>>>
\\ \nonumber
%<<<<<<<<<<<<<<<<<<<<<<<<<<<<<<<<<<<<<<<<<<<<<<<<<<<<<<<
\begin{picture}(200,45)(0,0)
\put(0,22){$\left(\! \frac{\sqrt{2} f_\pi \chi_1}{c_1}
\! -\! \frac{\gamma_{A_3}}{i}\!  \right)\! \gamma_5$}
\put(80,25){\oval(10,10)[l]} \put(72,23){$\bullet$}
\fermionleft{100}{30} \fermionright{80}{20}
\put(100,16){\framebox(10,18){}}
\fermionleft{130}{30} \fermionright{110}{20}
\put(130,25){\oval(10,10)[r]}
\put(130,27){$\bullet$} \put(135,35){$\Delta_2$}
\put(130,18){$\bullet$} \put(135,8){$\Delta_3$}
\end{picture}
\ea
From this expression, the two terms with $c_j$ are zero in the chiral
limit (as appropriate to this paper). The $c_j$'s diverge for low energies
in the chiral limit: they contain the pseudoscalar ladder pole, But in
this diagrams, the object to the right of the ladder contains a product of
2 $\Delta$'s, each of negative parity, the result carrying positive
parity. By the symmetry breaking pattern of the theory, no massless
scalar, pseudovector or tensor meson pole can make the ladder
divergent at low momentum. Therefore the terms with $c_j$ can be discarded and
 we have to consider only (substituting $\gamma_A=\frac{\slash P
}{i}\gamma_5$, good in the chiral limit):
\be
\begin{picture}(140,45)(0,0)
\put(-3,22){$\frac{i}{(\sqrt{2}f_\pi)^4} \slash P_3$}
\put(50,25){\oval(10,10)[l]} \put(42,23){$\bullet$}
\fermionleft{70}{30} \fermionright{50}{20}
\put(70,16){\framebox(10,18){}}
\fermionleft{100}{30} \fermionright{80}{20}
\put(100,25){\oval(10,10)[r]}
\put(100,27){$\bullet$} \put(105,35){$\Delta_1$}
\put(100,18){$\bullet$} \put(105,8){$\Delta_2$}
\put(130,22){$-$}
\end{picture}
%>>>>>>>>>>>>>>>>>>>>>>>>>>>>>>>>>>>>>>>>>>>>>>>>>><<
%<<<<<<<<<<<<<<<<<<<<<<<<<<<<<<<<<<<<<<<<<<<<<<<<<<<
\begin{picture}(140,45)(0,0)
\put(-3,22){$\frac{i}{(\sqrt{2} f_\pi)^4} \slash P_1 $}
\put(50,25){\oval(10,10)[l]} \put(42,23){$\bullet$}
\fermionleft{70}{30} \fermionright{50}{20}
\put(70,16){\framebox(10,18){}}
\fermionleft{100}{30} \fermionright{80}{20}
\put(100,25){\oval(10,10)[r]}
\put(100,27){$\bullet$} \put(105,35){$\Delta_2$}
\put(100,18){$\bullet$} \put(105,8){$\Delta_3$}
\end{picture}
\ee
Therefore our next problem is to evaluate diagrams such as:
\ba \label{VectortotwoDeltas}
\begin{picture}(90,45)(0,0)
\put(0,22){$\slash{P}_3$}
\put(20,25){\oval(10,10)[l]} \put(12,23){$\bullet$}
\fermionleft{40}{30} \fermionright{20}{20}
\put(20,35){k}
\put(15,7){$k\! \!-\! \! P_1\! \! - \! \! P_2$}
\put(40,16){\framebox(10,18){}}
\fermionleft{70}{30} \fermionright{50}{20}
\put(70,25){\oval(10,10)[r]}
\put(70,27){$\bullet$} \put(75,35){$\Delta_1$}
\put(70,18){$\bullet$} \put(75,8){$\Delta_2$}
\end{picture} \\ \nonumber
=3 d_3 P_1\cd P_3 \ P_1 \cd P_2 + 3 d_4 P_2 \cd P_3 \ P_1 \cd P_2
\ea
which again explicitly displays the color factor and where the new
constants $d_3$ and $d_4$ have to be calculated in a specific model.

To reduce the ladder in this diagram we could attempt to use the Vector Ward Identity
for the $\gamma_\mu$ on the left, but the momenta flowing in the adjoining
propagators would require this $\gamma_\mu$ to be contracted with
$-P_1-P_2 = P_3+P_4$ and not just with $P_3$. Or in the right part of the
diagram we could reuse our result
for the two-$\Delta$  ($\sqcup$) vertex from the previous section. But again
the momentum flow is not adequate. The solution to this impasse is to use
both ideas, but in a momentum expansion.The ladder in this diagram can be
substituted by its momentum expansion (\ref{ladderEXP}). Since there are
three powers of momentum already committed (one is the explicit $P_3$, the
other two need to be one in each $\Delta$) only one more power is needed.
Therefore we can use the ladder expansion to order one, and diagram
(\ref{VectortotwoDeltas}) can be rewritten as
\ba
  \begin{picture}(100,40)(0,0)
  \put(20,25){\oval(10,10)[l]} \put(12,23){$\bullet$}
  \fermionleft{40}{20} \fermionright{20}{30}
  \put(40,15){\framebox(10,20){}}
  \put(50,30){\line(1,0){5}} \put(55,27){x}
  \put(50,20){\line(1,0){5}} \put(55,17){x}
  \put(20,0){$O(P)$}
  \fermionright{70}{30} \fermionleft{90}{20}
  \put(90,25){\oval(10,10)[r]}
  \put(87,28){$\bullet$} \put(90,32){$\Delta_1$} \put(75,32){$q$}
  \put(87,18){$\bullet$} \put(90,13){$\Delta_2$} \put(65,8){$ ^{q\! +\! \! P_1 \! \! + \! \! P_2} $}
  \put(70,0){$O(P^3)$} \put(95,20){$ ^{q+P_1} $}
  \end{picture}
\\ \nonumber + \\ \nonumber
  \begin{picture}(110,40)(50,0)
  \put(20,25){\oval(10,10)[l]} \put(12,23){$\bullet$} \put(2,20){$\slash{P}_3$}
  \fermionleft{40}{20} \fermionright{20}{30}
  \put(30,8){$k$} \put(30,32){$k$}
  \put(40,15){\framebox(10,20){}}
  \put(50,30){\line(1,0){10}} \fermionleft{80}{20}
  \put(50,20){\line(1,0){10}} \fermionright{60}{30}
  \put(50,8){$(k'\! \! +\! \! P_1 \! \! +\! \! P_2)^{(1)}$}
  \put(60,32){$k'$}
  \put(20,-5){$O(P)$} \put(60,-5){$O(P)$} \put(100,-5){$O(P^2)$}
  \put(80,15){\framebox(10,20){}}
  \fermionright{90}{30} \fermionleft{110}{20}
  \put(110,25){\oval(10,10)[r]}
  \put(107,28){$\bullet$} \put(110,32){$\Delta_1$} \put(95,32){$q$}
  \put(107,18){$\bullet$} \put(110,13){$\Delta_2$}
  \end{picture}
\ea

Now we can use the vector Ward-Takahashi Identity, 
which is satisfied to $O(P^{(1)})$, on the
first diagram and on the left ladder of the second diagram,
allowing us to substitute $$\slash{P}_3 \longrightarrow i
S^{-1}(q)-i S^{-1}(q-P_3) \ .$$
The matrix object with a ladder and two
deltas, to $O(P^{(2)})$, which appears on the second diagram, is
just $\sqcup$ as defined in (\ref{Definesqcup}). Now it is
straightforward to show that (\ref{VectortotwoDeltas}) is equal to
\be \label{closedelta3}
(2 k\cd P_3(A' \slash{k} -B')+A\! \! \slash{P}_3) \cd \left(
\begin{picture}(65,25)(0,0)
\put(20,5){\oval(10,10)[l]} \fermionleft{40}{0}
\fermionright{20}{10} \put(40,5){\oval(10,10)[r]}
\put(40,8){$\bullet$} \put(40,-2){$\bullet$}
\put(47,9){$\Delta_1$} \put(47,-3){$\Delta_2$}
\put(20,12){$q$} \put(10,-12){$^{(q\! +\! P_1\! +\! P_2)}$}
\end{picture}
+
\begin{picture}(65,25)(0,0)
\put(20,5){\oval(10,10)[l]} \fermionleft{40}{0}
\fermionright{20}{10} \put(40,5){\oval(10,10)[r]}
\put(20,12){$q$} \put(10,-12){$^{(q\! +\! P_1\! +\! P_2)^{(1)}}$}
\put(43,3){$\bullet$} \put(47,3){$i \sqcup$}
\end{picture}
\right)
\ee
where the diagrams have to be evaluated to order $(P^{(3)})$ since an 
explicit power of $P$ has already been used. The left diagram in particular
gives rise to four simple sub-diagrams since two powers of $P$ are 
committed in the $\Delta$'s, but the other power can be distributed  
alternatively between these $\Delta$'s or the two propagators which
carry a power of $P$.

These diagrams can all be evaluated easily as a simple loop integral
in the computer to obtain $d_3$, $d_4$. The contribution from 
(\ref{OneGThreeD}) is finally
\ba \label{contribution2} \nonumber
\frac{3}{(\sqrt{2}f_\pi)^4}(d_3(P_1\cd P_3 \ P_1 \cd P_2 - P_1 \cd P_2 \ P_2 \cd P_3)
\\ \nonumber
+d_4(P_2 \cd P_3 \ P_1\cd P_2 -P_1 \cd P_3 \ P_2 \cd P_3) + {\rm permutations}) \\
\ea

%%%%%%%%%%%%%%%%%%%%%%%%%%%%%%%%%%%%%%%%%%%%%%%%%%%%%%%%%%%%%%%%%%%%%%%
%%%%%%%%%%%%%%%%%%%%%%%%%%%%%%%%%%%%%%%%%%%%%%%%%%%%%%%%%%%%%%%%%%%%%%%
%%%%%%%%%%%%%%%%%%%%%%%%%%%%%%%%%%%%%%%%%%%%%%%%%%%%%%%%%%%%%%%%%%%%%%%
\subsection{Contribution with $\Delta \Gamma \Gamma \Gamma$.}
%%%%%%%%%%%%%%%%%%%%%%%%%%%%%%%%%%%%%%%%%%%%%%%%%%%%%%%%%%%%%%%%%%%%%%
The contribution
\be \label{OneDThreeG}
%>>>>>>>>>>>>>>>>>>>>>>>>>>>>>>>>>>>>>>>>>>>>>>
\begin{picture}(50,65)(0,0)
%%%%%%%%%%%%%%%%%%%%%%%%%%%%%% box with two gammas
\put(10,13){\begin{picture}(20,40)(0,0)
\put(5,0){\line(0,1){5}}
\put(5,15){\vector(0,-1){10}}
\put(15,15){\line(0,-1){5}}
\put(15,0){\vector(0,1){10}}
\put(0,15){\framebox(20,10){}}
\put(5,25){\line(0,1){5}}
\put(5,40){\vector(0,-1){10}}
\put(15,40){\line(0,-1){5}}
\put(15,25){\vector(0,1){10}}
\end{picture}}
\put(10,8){\begin{picture}(20,40)(0,0)
\put(10,40){\oval(10,26)[t]}
\put(2,42){$\bullet$}
\put(-9,48){$\Gamma_4$}
\put(13,42){$\bullet$}
\put(18,48){$\Delta_1$}
\put(10,53){\vector(-1,0){2}}
\end{picture}}
\put(10,-27){\begin{picture}(20,40)(0,0)
\put(10,45){\oval(10,26)[b]}
\put(2,38){$\bullet$}
\put(-9,34){$\Gamma_3$}
\put(13,38){$\bullet$}
\put(18,34){$\Gamma_2$}
\put(10,32){\vector(1,0){2}}
\end{picture}}
\put(40,30){+}
\end{picture}
%<<<<<<<<<<<<<<<<<<<<<<<<<<<<<<<<<<<<<<<<<<<<<<<<<<<
%>>>>>>>>>>>>>>>>>>>>>>>>>>>>>>>>>>>>>>>>>>>>>>
\begin{picture}(100,50)(0,0) \put(0,11){
\begin{picture}(100,40)(0,0)
%%%%%%%%%%%%%%%%%%%%%%%%%%%%%% box with two gammas
\put(25,10){\begin{picture}(40,20)(0,0)
\put(0,15){\line(1,0){5}}
\put(0,5){\vector(1,0){10}}
\put(15,15){\vector(-1,0){10}}
\put(15,5){\line(-1,0){5}}
\put(15,0){\framebox(10,20){}}
\put(25,15){\line(1,0){5}}
\put(25,5){\vector(1,0){10}}
\put(40,15){\vector(-1,0){10}}
\put(40,5){\line(-1,0){5}}
\end{picture}}
\put(25,10){\begin{picture}(40,20)(0,0)
\put(40,10){\oval(30,10)[r]}
\put(42,2){$\bullet$}
\put(48,-4){$\Gamma_2$}
\put(42,13){$\bullet$}
\put(48,20){$\Delta_1$}
\put(55,10){\vector(0,1){2}}
\end{picture}}
\put(-20,10){\begin{picture}(40,20)(0,0)
\put(45,10){\oval(30,10)[l]}
\put(38,2){$\bullet$}
\put(34,-4){$\Gamma_3$}
\put(38,13){$\bullet$}
\put(34,20){$\Gamma_4$}
\put(30,10){\vector(0,-1){2}}
\end{picture}}
\end{picture} } %end put
\put(95,30){$-$}
\end{picture}
%<<<<<<<<<<<<<<<<<<<<<<<<<<<<<<<<<<<<<<<<<<<<<<<<<<<
%>>>>>>>>>>>>>>>>>>>>>>>>>>>>>>>>>>>>>>>>>>>>>>>>>>>
\begin{picture}(65,50)(0,0) \put(0,11){
\begin{picture}(65,40)(0,0)
%%%%%%%%%%%%%%%%%%%%%%%%%%%%%% four Gammas
\put(25,10){\begin{picture}(40,20)(0,0)
\put(0,15){\line(1,0){5}}
\put(0,5){\vector(1,0){10}}
\put(15,15){\vector(-1,0){10}}
\put(15,5){\line(-1,0){5}}
\end{picture}}
\put(0,10){\begin{picture}(40,20)(0,0)
\put(40,10){\oval(30,10)[r]}
\put(42,2){$\bullet$}
\put(48,-4){$\Gamma_2$}
\put(42,13){$\bullet$}
\put(48,20){$\Delta_1$}
\put(55,10){\vector(0,1){2}}
\end{picture}}
\put(-20,10){\begin{picture}(40,20)(0,0)
\put(45,10){\oval(30,10)[l]}
\put(38,2){$\bullet$}
\put(34,-4){$\Gamma_3$}
\put(38,13){$\bullet$}
\put(34,20){$\Gamma_4$}
\put(30,10){\vector(0,-1){2}}
\end{picture}}
\end{picture} } %end put
\end{picture}
%<<<<<<<<<<<<<<<<<<<<<<<<<<<<<<<<<<<<<<<<<<<<<<<<<<<
\ee
can be reduced as follows:
1) apply (\ref{expandGamma}) and (\ref{Laddersaturation}) to the $\Delta_1$ in
the first term, and (\ref{GAvertexgraph}) to the $\Gamma_3$
  in the second diagram 
to obtain an expression similar to (\ref{OneGThreeD2}) in which $\Gamma_4$ is 
isolated. 2) Employ (\ref{upperWI2}) to eliminate $\Gamma_4$. 3) Repeat the 
operation to eliminate $\Gamma_2$. Obtain
\ba
\begin{picture}(100,45)(0,0)
\put(0,22){$\slash{P}_1$}
\put(20,25){\oval(10,10)[l]} \put(12,23){$\bullet$}
\fermionleft{40}{30} \fermionright{20}{20}
\put(40,16){\framebox(10,18){}}
\fermionleft{70}{30} \fermionright{50}{20}
\put(70,25){\oval(10,10)[r]}
\put(73,23){$\bullet$} \put(75,22){$\slash{P}_3$}
\put(25,32){$q$} \put(17,5){$^{q\! + \! P_1 \! +\! P_4}$}
\put(90,23){+}
\end{picture}
\nonumber \\ 
\begin{picture}(100,45)(0,0)
\put(0,22){$\slash{P}_1$}
\put(20,25){\oval(10,10)[l]} \put(12,23){$\bullet$}
\fermionleft{40}{30} \fermionright{20}{20}
\put(40,16){\framebox(10,18){}}
\fermionleft{70}{30} \fermionright{50}{20}
\put(70,25){\oval(10,10)[r]}
\put(73,23){$\bullet$} \put(75,22){$\slash{P}_3$}
\put(25,32){$q$} \put(17,5){$^{q\! + \! P_1 \! +\! P_2}$}
\end{picture}
 \nonumber \\ 
+ {\rm permutations} \ .
\ea
In this expression, two explicit powers of $P$ are present, namely $P_1$ 
and $P_3$ in the vertices. The other two powers have to be produced from
a propagator expansion. 
This is the third (and last) different diagram type that we need to parameterize:
\ba \label{Diag3}
\begin{picture}(100,45)(0,0)
\put(0,22){$\slash{P}_1$}
\put(20,25){\oval(10,10)[l]} \put(12,23){$\bullet$}
\fermionleft{40}{30} \fermionright{20}{20}
\put(40,16){\framebox(10,18){}}
\fermionleft{70}{30} \fermionright{50}{20}
\put(70,25){\oval(10,10)[r]}
\put(73,23){$\bullet$} \put(75,22){$\slash{P}_3$}
\put(25,32){$q$} \put(17,5){$^{q\! + \! P_1 \! +\! P_4}$}
\end{picture} \\ \nonumber
= 3 d_5 P_1 \cd P_3 \ P_1 \cd P_4 + 3 d_6 P_3 \cd P_4 \ P_1 \cd P_4 
\ea
Next we show how to calculate $d_5$, $d_6$.
Since there is a ladder which contains powers of $P$,
we need to recall the ladder expansion to second order in (\ref{ladderEXP}).
Eliminating the loose rung with the help of (\ref{looserungkill}),
and employing the vector Ward Identity to generate a vertex 
$$V(P)=2 q\cd P(A'(q)\slash{q}
-B'(q))+A(q)\slash{P} \ ,$$ we can show
\begin{widetext}
\be \label{Diag3expanded}
(\ref{Diag3})= 
\begin{picture}(100,30)(0,0)
\put(30,20){\oval(10,10)[l]} \put(50,20){\oval(10,10)[r]}
\fermionleft{50}{25} \fermionright{30}{15}
\put(23,18){$\bullet$} \put(0,16){$^{V(P_1)}$}
\put(53,18){$\bullet$} \put(60,16){$^{V(P_3)}$}
\put(40,27){$q$} \put(25,5){$^{(q\! +\! P_1\! +\! P_4)^{(2)}}$}
\end{picture}
\begin{picture}(150,30)(0,0) \put(-12,17){+}
\put(30,20){\oval(10,10)[l]} \put(80,20){\oval(10,10)[r]}
\fermionleft{50}{25} \fermionright{30}{15}
\put(50,10){\framebox(10,20){}}
\fermionleft{80}{25} \fermionright{60}{15}
\put(23,18){$\bullet$} \put(0,16){$^{V(P_1)}$}
\put(83,18){$\bullet$} \put(90,16){$^{V(P_3)}$}
\put(40,27){$q$}  \put(5,0){$^{(q\! +\! P_1\! +\! P_4)^{(1)}}$}
\put(70,27){$k$}  \put(60,0){$^{(k\! +\! P_1\! +\! P_4)^{(1)}}$}
\end{picture}
\begin{picture}(150,30)(0,0) \put(-12,17){-}
\put(30,20){\oval(10,10)[l]} \put(70,20){\oval(10,10)[r]}
\fermionleft{50}{25} \fermionright{30}{15}
\fermionleft{70}{25} \fermionright{50}{15}
\put(23,18){$\bullet$} \put(0,16){$^{V(P_1)}$}
\put(73,18){$\bullet$} \put(80,16){$^{V(P_3)}$}
\put(50,27){$q$} \put(50,13){$\bullet$}
\put(5,0){$^{(q\! +\! P_1\! +\! P_4)^{(1)}}$}
\put(45,14){$^{S(q)^{-\! 1}}$}
\put(55,0){$^{(q\! +\! P_1\! +\! P_4)^{(1)}}$}
\end{picture}
\ee
\end{widetext}

The first and third diagrams are again straightforward traces and
integrals. Only the middle diagram contains a ladder. We can interpret
this ladder as ``dressing'' either of the vertices, and write immediately
an integral equation for one of them. Taking for example
\be \label{Definevee}
\begin{picture}(300,30)(0,0)  \put(120,20){\oval(10,10)[l]} 
\fermionleft{140}{25} \fermionright{120}{15}
\put(0,16){$^{[2 kP_1(A'(k)\slash{k}-B'(k))+A(k)\slash{P}_1]}$}
\put(113,17){$\bullet$}
\put(130,27){$k$} 
\put(110,0){$^{(k\! +\! P_1\! +\! P_4)^{(1)}}$}
\put(140,11){\framebox(10,18){}}
\fermionleft{170}{25} 
\fermionright{150}{15}
\put(168,13){x} \put(168,23){x}
\put(190,15){$=$} \put(210,15){$\vee$}
\end{picture}
\ee
the function vertex $\vee$ so defined satisfies a linear inhomogeneous
integral equation (the first argument is the momentum entering the diagram
through the vertex, the second is the relative and the third the total 
momentum between the fermion lines at the vertex):
\ba \label{eqnforvee}
\vee(P_1,k+\frac{P_1+P_4}{2},P_1+P_4)= \nonumber \\ \nonumber
[2k P_1(A'(k)\! \slash{k}\! -\! B'(k))+A(k)\! \slash{P}_1] 
S(k\! +\! P_1\! +\! P_4)^{(1)}S(k)^{-1} \\ \nonumber \\
\begin{picture}(150,30)(0,0) 
\put(0,17){$+ \ \vee$}
\put(23,17){$\bullet$}
\put(30,20){\oval(10,10)[l]} 
\fermionleft{50}{25} \fermionright{30}{15}
\put(38,28){$q$} \put(38,5){$q$}
\multiput(42,13)(0,2){6}{$\cdot$} \put(48,13){x} \put(48,23){x}
\end{picture}
\ea

As is evident, $\vee$ admits an expansion up to second order in momentum
identical to (\ref{expandsqcup}) in terms of a new set of functions
$\vee_0(k^2)$, ... ,$\vee_9(k^2)$. The integral system of equations 
(\ref{eqnforvee})
is very similar to (\ref{eqnforsqcup}), the only difference being the
inhomogeneous term. Therefore the linear projections in (\ref{Linearsqcup})
still apply, and both systems can be solved with basically the same iterative
computer code. 

Finally the middle diagram in (\ref{Diag3expanded}) can be closed to read
\be \label{closedelta2}
\begin{picture}(150,30)(0,0) 
\put(30,20){\oval(10,10)[l]} \put(70,20){\oval(10,10)[r]}
\fermionleft{50}{25} \fermionright{30}{15}
\fermionleft{70}{25} \fermionright{50}{15}
\put(23,18){$\bullet$} \put(0,16){$\vee$}
\put(73,18){$\bullet$} \put(80,16){$^{V(P3)}$}
\put(50,27){$q$} 
\put(25,0){$^{(q\! +\! P_1\! +\! P_4)^{(1)}}$}
\end{picture}
\ee
which is easy to evaluate with the help of a symbolic manipulation program.
Finally we give the expression for (\ref{OneDThreeG}) in terms of the $d$'s:
\ba \label{contribution3}
3 d_5 P_1\cd P_3(P_1\cd P_4+P_1 \cd P_2) + \\ \nonumber
3 d_6 (P_1 \cd P_2 P_2 \cd P_3 + P_1 \cd P_4 P_3\cd P_4) + {\rm permutations }\ .
\ea

%%%%%%%%%%%%%%%%%%%%%%%%%%%%%%%%%%%%%%%%%%%%%%%%%%%%%%%%%%%%%%%%%%%%%%%
%%%%%%%%%%%%%%%%%%%%%%%%%%%%%%%%%%%%%%%%%%%%%%%%%%%%%%%%%%%%%%%%%%%%%%%
%%%%%%%%%%%%%%%%%%%%%%%%%%%%%%%%%%%%%%%%%%%%%%%%%%%%%%%%%%%%%%%%%%%%%%%
\subsection{Contribution with $\Delta \Delta \Gamma \Gamma$.}
%%%%%%%%%%%%%%%%%%%%%%%%%%%%%%%%%%%%%%%%%%%%%%%%%%%%%%%%%%%%%%%%%%%%%%
With two $\Delta$ corrections, there are two topologically distinct
diagrams that can contribute. They are different because while reading
around the fermion loop one can
find the external legs in the order$\Gamma \Gamma \Delta \Delta$ or in
the order $\Gamma \Delta \Gamma \Delta$ .
We start by the first term, namely
\be
%>>>>>>>>>>>>>>>>>>>>>>>>>>>>>>>>>>>>>>>>>>>>>>
\begin{picture}(50,65)(0,0)
%%%%%%%%%%%%%%%%%%%%%%%%%%%%%% box with two gammas
\put(10,13){\begin{picture}(20,40)(0,0)
\put(5,0){\line(0,1){5}}
\put(5,15){\vector(0,-1){10}}
\put(15,15){\line(0,-1){5}}
\put(15,0){\vector(0,1){10}}
\put(0,15){\framebox(20,10){}}
\put(5,25){\line(0,1){5}}
\put(5,40){\vector(0,-1){10}}
\put(15,40){\line(0,-1){5}}
\put(15,25){\vector(0,1){10}}
\end{picture}}
\put(10,8){\begin{picture}(20,40)(0,0)
\put(10,40){\oval(10,26)[t]}
\put(2,42){$\bullet$}
\put(-9,48){$\frac{\Gamma_4}{i}$}
\put(13,42){$\bullet$}
\put(18,48){$\Delta_1$}
\put(10,53){\vector(-1,0){2}}
\end{picture}}
\put(10,-27){\begin{picture}(20,40)(0,0)
\put(10,45){\oval(10,26)[b]}
\put(2,38){$\bullet$}
\put(-9,34){$\frac{\Gamma_3}{i}$}
\put(13,38){$\bullet$}
\put(18,34){$\Delta_2$}
\put(10,32){\vector(1,0){2}}
\end{picture}}
\put(40,30){+}
\end{picture}
%<<<<<<<<<<<<<<<<<<<<<<<<<<<<<<<<<<<<<<<<<<<<<<<<<<<
%>>>>>>>>>>>>>>>>>>>>>>>>>>>>>>>>>>>>>>>>>>>>>>
\begin{picture}(100,50)(0,0) \put(0,11){
\begin{picture}(100,40)(0,0)
%%%%%%%%%%%%%%%%%%%%%%%%%%%%%% box with two gammas
\put(25,10){\begin{picture}(40,20)(0,0)
\put(0,15){\line(1,0){5}}
\put(0,5){\vector(1,0){10}}
\put(15,15){\vector(-1,0){10}}
\put(15,5){\line(-1,0){5}}
\put(15,0){\framebox(10,20){}}
\put(25,15){\line(1,0){5}}
\put(25,5){\vector(1,0){10}}
\put(40,15){\vector(-1,0){10}}
\put(40,5){\line(-1,0){5}}
\end{picture}}
\put(25,10){\begin{picture}(40,20)(0,0)
\put(40,10){\oval(30,10)[r]}
\put(42,2){$\bullet$}
\put(48,-4){$\Delta_2$}
\put(42,13){$\bullet$}
\put(48,20){$\Delta_1$}
\put(55,10){\vector(0,1){2}}
\end{picture}}
\put(-20,10){\begin{picture}(40,20)(0,0)
\put(45,10){\oval(30,10)[l]}
\put(38,2){$\bullet$}
\put(34,-4){$\frac{\Gamma_3}{i}$}
\put(38,13){$\bullet$}
\put(34,20){$\frac{\Gamma_4}{i}$}
\put(30,10){\vector(0,-1){2}}
\end{picture}}
\end{picture} } %end put
\put(95,30){$-$}
\end{picture}
%<<<<<<<<<<<<<<<<<<<<<<<<<<<<<<<<<<<<<<<<<<<<<<<<<<<
%>>>>>>>>>>>>>>>>>>>>>>>>>>>>>>>>>>>>>>>>>>>>>>>>>>>
\begin{picture}(65,50)(0,0) \put(0,11){
\begin{picture}(65,40)(0,0)
%%%%%%%%%%%%%%%%%%%%%%%%%%%%%% four Gammas
\put(25,10){\begin{picture}(40,20)(0,0)
\put(0,15){\line(1,0){5}}
\put(0,5){\vector(1,0){10}}
\put(15,15){\vector(-1,0){10}}
\put(15,5){\line(-1,0){5}}
\end{picture}}
\put(0,10){\begin{picture}(40,20)(0,0)
\put(40,10){\oval(30,10)[r]}
\put(42,2){$\bullet$}
\put(48,-4){$\Delta_2$}
\put(42,13){$\bullet$}
\put(48,20){$\Delta_1$}
\put(55,10){\vector(0,1){2}}
\end{picture}}
\put(-20,10){\begin{picture}(40,20)(0,0)
\put(45,10){\oval(30,10)[l]}
\put(38,2){$\bullet$}
\put(34,-4){$\frac{\Gamma_3}{i}$}
\put(38,13){$\bullet$}
\put(34,20){$\frac{\Gamma_4}{i}$}
\put(30,10){\vector(0,-1){2}}
\end{picture}}
\end{picture} } %end put
\end{picture}
%<<<<<<<<<<<<<<<<<<<<<<<<<<<<<<<<<<<<<<<<<<<<<<<<<<<
\ee
(and the corresponding three more permutations).
Again by applying (\ref{Laddersaturation}) to the $\Delta_1$ in the first
diagram  and (\ref{GAvertex}) to the $\Gamma_3$ in the second,
then using (\ref{GAvertex2}) to simplify the remaining $\Gamma_4$,
neglecting the positive parity ladders when they are divided by a $c_j$
containing the pion pole, re-absorbing the ladders and simplifying, one
obtains:
\ba
\begin{picture}(150,45)(0,0)
\put(0,22){$\frac{\slash{P}_3}{i}$}
\put(40,25){\oval(10,10)[l]} \put(32,23){$\bullet$}
\fermionleft{60}{30} \fermionright{40}{20}
\put(50,32){$q$}
\put(60,16){\framebox(10,18){}}
\fermionleft{90}{30} \fermionright{70}{20}
\put(90,25){\oval(10,10)[r]}
\put(90,27){$\bullet$} \put(95,35){$\Delta_1$}
\put(90,18){$\bullet$} \put(95,8){$\Delta_2$}
\put(120,22){$+$}
\put(50,32){$q$} \put(30,5){$^{(q\! + \! P_3 \! + \! P_4)}$}
\end{picture}
%>>>>>>>>>>>>>>>>>>>>>>>>>>>>>>>>>>>>>>>>>>>>>>>>>>>>>>>>
\\ \nonumber
%<<<<<<<<<<<<<<<<<<<<<<<<<<<<<<<<<<<<<<<<<<<<<<<<<<<<<<<
\begin{picture}(150,45)(0,0)
\put(0,22){$ \slash{P}_1$}
\put(50,32){$q$}
\put(40,25){\oval(10,10)[l]} \put(32,23){$\bullet$}
\fermionleft{60}{30} \fermionright{40}{20}
\put(50,32){$q$} \put(30,5){$^{(q\! + \! P_1 \! + \! P_4)}$}
\put(60,16){\framebox(10,18){}}
\fermionleft{90}{30} \fermionright{70}{20}
\put(90,25){\oval(10,10)[r]}
\put(95,23){$\bullet$} \put(105,23){$\slash{P}_2$}
\end{picture}
\\ \nonumber
+ {\rm permutations}
\ea
which can be written down immediately in terms of the $d's$ defined in eqns. 
(\ref{VectortotwoDeltas}), (\ref{Diag3}) as 
\ba \label{contribution4}
3[ -d_3 P_1 \cd P_3 P_1 \cd P_2 - d_4 P_2 \cd P_3 P_1 \cd P_2 \\ \nonumber
+d_5 P_1 \cd P_2 P_1 \cd P_4 +d_6 P_2 \cd  P_4 P_1 \cd P_4 + {\rm permutations} 
] \ .
\ea

%%%%%%%%%%%%%%%%%%%%%%%%%%%%%%%%%%%%%%%%%%%%%%%%%%%%%%%%%%%%%%%%%%%%%%%
\subsection{$\Delta \Gamma \Delta \Gamma$ }
%%%%%%%%%%%%%%%%%%%%%%%%%%%%%%%%%%%%%%%%%%%%%%%%%%%%%%%%%%%%%%%%%%%%%%%
Two permutations contribute: $\Delta_1 \Gamma_2 \Delta_3 \Gamma_4$ and
$\Gamma_1 \Delta_2 \Gamma_3 \Delta_4$. The reduction is in all analogous
to the previous ones, yielding a contribution
\ba \label{contribution5}
\frac{-3}{(\sqrt{2} f_\pi)^4} [d_5(P_1\cd P_3 \ P_1 \cd P_4 +P_1\cd P_3
\ P_3\cd P_4 + {\rm permutation}) \nonumber \\ \nonumber
+d_6(P_3\cd P_4 \ P_1 \cd P_4+P_1\cd P_4 \ P_3 \cd P_4 + {\rm permutation})] \\
\ea

%%%%%%%%%%%%%%%%%%%%%%%%%%%%%%%%%%%%%%%%%%%%%%%%%%%%%%%%%%%%%%%%%%%%
%%%%%%%%%%%%%%%%%%%%%%%%%%%%%%%%%%%%%%%%%%%%%%%%%%%%%%%%%%%%%%%%%%%%
\subsection{$4-\Gamma$ contribution.}
%%%%%%%%%%%%%%%%%%%%%%%%%%%%%%%%%%%%%%%%%%%%%%%%%%%%%%%%%%%%%%%%%%%%
The last piece stems from the term with four powers of $\Gamma$. By repeated
use of (\ref{upperWI2}) it can be shown to contribute
\ba \label{contribution6}
\frac{-3}{(\sqrt{2} f_\pi)^4} [d_5(P_1\cd P_3 \ P_3 \cd P_4 +P_1\cd P_3
\ P_2\cd P_3)  \nonumber \\ 
+d_6(P_1\cd P_4 \ P_3 \cd P_4+P_1\cd P_2 \ P_2 \cd P_3) ] \ .
\ea
 (Here the cyclicity of (\ref{PIPISCATT}) can be recovered by using 
$\sum_i P_i=0$).

Combining the results from subsections A through F and summing all 
permutations  (\ref{permutations}) with the isospin factors
(\ref{isospinfirst}, \ref{isospinsecond}), the standard amplitude
$A(s,t,u)$, takes a form identical to (\ref{standardamplitude})
and  we deduce 
\ba 
l_1&=& { 3 \over 32}(-2 d_1 +  d_2 - 6 d_3 + 2 d_4 + 5 d_5 +  d_6)\ 
\nonumber \\
l_2&=& -{3 \over 16} (d_2 - 2d_3 + 2d_4 +d_5 -d_6)
\ . 
\ea 
Therefore, to obtain the $l$'s numerically, one must: 
\begin{enumerate}
\item Solve the Schwinger-Dyson equations for the propagator, equation
(\ref{SD}).
\item Employ the obtained $A$, $B$ functions as input to the Bethe-Salpeter
equations (\ref{BS}) and solve them.
\item Then use the obtained $F_0$, $G_0$, ... as input for equations
(\ref{eqnforsqcup}) and (\ref{eqnforvee}) to obtain $\sqcup$ and $\vee$.
\item Perform the integrals (\ref{closedelta4}), (\ref{closedelta3}),
(\ref{closedelta2}).
\item Assemble (\ref{defined1d2}), (\ref{contribution2}), (\ref{contribution3}),
(\ref{contribution4}), (\ref{contribution5}), (\ref{contribution6}).
\end{enumerate}
In this external momentum expansion all integrals and integral equations 
are functions of internal variables only of quark momenta 
$q^2$, $k^2$, $k\cd q$. The diagrams could alternatively be evaluated on
the lattice.

%%%%%%%%%%%%%%%%%%%%%%%%%%%%%%%%%%%%%%%%%%%%%%%%%%%%%%%%%%%%%%%%%%%%%
\section{Model Evaluations.} \label{modelitos}
%%%%%%%%%%%%%%%%%%%%%%%%%%%%%%%%%%%%%%%%%%%%%%%%%%%%%%%%%%%%%%%%%%%%%
We would like to provide simple model evaluations of all these
 calculations, well aware of model limitations and that a thorough
phenomenological analysis can only be carried out with more
sophisticated interactions such as employed in \cite{Roberts,Maris}.
We employ two Feynman gauge models featuring an interaction
\be \label{Feynman}
V K(q) V = \gamma_\mu  K(q)  \gamma^\mu \ .
\ee

This simple choice of a vector-vector interaction (as opposed
to the more popular Landau gauge transverse tensor kernel) simplifies
the Gamma matrix traceology (in this calculation carried out with the
help of two independent computer codes, one written in MATHEMATICA and
one in FORM \cite{Vermasseren}) so that the standard Llewelyn-Smith
BS wavefunction for the pion \cite{Maris} reduces to
\ba \label{chiexpansion} \nonumber
\chi(P,k)&=& \!  \gamma_5\left( E(P,k)+F(P,k) \!
\slash{P}+G(P,k)\slash{k}k\cd P \right) \\ \nonumber
&=&\gamma_5 \left(  E0(k^2)+\frac{(kp)^2}{2}E2(k^2)+F0(k^2) \slash{P} \right. \\ 
 & & + \left. G0(k^2)\slash{k}  k\cd P + \cdot \cdot \cdot \right)
\ea 
where the
standard $H$ function decouples from the rest of the system and
therefore we further ignore, and the momentum expansion shown is
complete up to second order for a symmetric momentum routing
$\chi$ (that is, the fermion lines out of $\chi$ carry $k+P/2$
and $k-P/2$). The $E0$ term is determined by chiral symmetry to be
$\frac{\Gamma_A(P=0,k)}{\sqrt{2}if_\pi}$. The rest of the
expansion constructs the function $\Delta(P,k)$. 
The same power series (\ref{chiexpansion})
can be written down for  the axial vertex,
\ba&
\Gamma_A(P,q)= [(A(q-P/2)(\slash{q}-\frac{\slash{P}}{2})- \\ \nonumber &
A(q+P/2)(\slash{q}+\frac{\slash{P}}{2})
-(B(q-P/2)+B(q+P/2))]\frac{\gamma_5}{i}
\ea
expanding in powers of $P$, and up to a normalization we recover
the equivalent to (\ref{chiexpansion})
\ba \label{Gammaexpansion}
E0_A= 2 B(q^2) \\ \nonumber
F0_A=-A(q^2) \\ \nonumber
G0_A=-2 A'(q^2) \\ \nonumber
E2_A= 2 B''(q^2) \\ \nonumber
... 
\ea
Subtracting (\ref{Gammaexpansion}) from (\ref{chiexpansion})
we obtain some new functions of $q^2$ which provide the needed
expansions for $\Delta$:
\begin{widetext}
\ba
\label{Deltaexpansion}
 \Delta^{(1)}(q+P_1/2,P_1)=\Delta^{(1)}(q,P_1)=\ov{F0}(q^2)
\slash{P}_1 +\ov{G0}(q^2) q\cd P_1 \slash{q} \\ \nonumber
\Delta^{(2)}(q+P_1/2,P_1)= E^2(q^2) \frac{(q\cd P_1)^2}{2} +
\ov{F0}'(q^2) \slash{P}_1 q.P_1 + \ov{G0}'(q^2)\slash{q} (q\cd
 P_1)^2
 \ea
\end{widetext}
 (valid for symmetric momentum routing when $E1$, $F1$, $G1$ all
 vanish).
Further, with the $\gamma_\mu \gamma^\mu$ kernel another trace-related 
simplification occurs in (\ref{eqnforsqcup}) and (\ref{eqnforvee}),
and the functions $\sqcup_6$, $\sqcup_7$, $\sqcup_8$, $\vee_6$,
$\vee_7$, $\vee_8$  equal the inhomogeneous term
in their respective equations, the homogeneous (integral) parts of
the equations being zero.

All that remains is to consider some specific form for $K(q^2)$.
We will look at two models whose euclidean angular integrals can be
done analytically, leaving only one-dimensional integral equations
to solve numerically.

The first is a simple Gaussian kernel (whose Euclidean angular
integrals are Bessel functions \cite{Alkofer})
\be 
K(q)=g^2 exp(-q^2/\Lambda^2)
\ee
where $g$ provides the coupling strength and $\Lambda$ the
scale of the interaction.
(the results are sketched in table \ref{gaussianmodeltable}).

\begin{table}[h]%[H] add [H] placement to break table across pages
\caption{\label{gaussianmodeltable}
Results for the toy gaussian model in the chiral limit. Dimensionful magnitudes are in MeV.}
\begin{ruledtabular}
\begin{tabular}{cc|ccccc}
 $\Lambda$ & g  & $M(k^2=0)$ &  $-\la \ov{\Psi}\Psi \ra^{1/3}$ &   
$f_\pi$ & $l_1$ & $l_2$       \\ \hline
500 & 6.3 & 385 & 222 & 76 & -0.02 & 0.066 \\
600 & 6.3 & 462 & 267 & 91 & -0.02 & 0.060 \\
500 & 6.5 & 468 & 237 & 83 & -0.018 & 0.062 \\
800 & 5.5 & 125 & 213 & 51 & -0.019 & 0.080 \\
\end{tabular}
\end{ruledtabular}
\end{table}

The second is a rational kernel
\be
 K(q)=g^2 [ 1/(q^2-\lambda^2) -1/(q^2-\Lambda^2)] 
\ee

\begin{table}[h]%[H] add [H] placement to break table across pages
\caption{\label{rationalmodeltable}
Results for the toy rational model in the chiral limit. Dimensionful magnitudes are in MeV.}
\begin{ruledtabular}
\begin{tabular}{ccc|ccccc}
 $\lambda$ &$\Lambda$ &g  & $M(k^2=0)$ &  $-\la \ov{\Psi}\Psi \ra^{1/3}$ &   
$f_\pi$ & $l_1$ & $l_2$       \\ \hline
250 & 300 & 13.0 & 559 & 242 & 82 & -0.011 & 0.11 \\
300 & 350 & 13.0 & 471 & 240 & 77 & -0.015 & 0.11 \\
500 & 550 & 14.2 & 269 & 247 & 68 & -0.015 & 0.12 \\
700 & 770 & 13.4 & 155 & 250 & 56 & -0.012 & 0.12 \\
\end{tabular}
\end{ruledtabular}
\end{table}

where g, $\lambda$ and $\Lambda$ represent some quark model parameters
fit to provide a good condensate and constituent quark mass via
the Schwinger-Dyson equation.
 The euclidean angular integrals
are straight-forward using relations such as
$$
\int_{-1}^1 \frac{\sqrt{1-x^2}}{a-x}= \pi (a-\sqrt{a^2-1})\ ,
$$
$$
\int_{-1}^1 \frac{x \sqrt{1-x^2}}{a-x}= \pi (a^2-1/2-a\sqrt{a^2-1}) \ .
$$
Hence all integral equations are one-dimensional in momentum space.

When the dimensionful  parameters, of order the strong interaction scale,
 are close in value, and for large enough $g$, 
then the potential is infrared enhanced supporting chiral symmetry breaking. 
Notice that the Rainbow-Ladder fermion loop diagrams constructed with these
interactions are finite due to the exponential or $1/q^4$ 
high energy behavior and no renormalization program is needed. This is 
associated to a gluon propagator scale, which determines the scale of 
the Bethe-Salpeter wavefunctions, which in turn regulate the meson loops
that would appear in chiral perturbation theory. We are currently 
investigating this issues.

%%%%%%%%%%%%%%%%%%%%%%%%%%%%%%%%%%%%%%%%%%%%%%%%%%%%%%%%%%%%%%%%%%%%%
\section{Results and Discussion}
%%%%%%%%%%%%%%%%%%%%%%%%%%%%%%%%%%%%%%%%%%%%%%%%%%%%%%%%%%%%%%%%%%%%%
We have worked out the pion-pion scattering amplitude at low energy
in the chiral limit to $O(P^4)$ from a general microscopic quark
Schwinger-Dyson approach.  Upon comparison with the chiral lagrangian
formalism, the main result of this paper is a pair of relations that would
in principle allow to directly evaluate the coefficients $l_1$, $l_2$ in
any specific model (i.e. after choosing the Lorentz structure of the
quark-antiquark interaction, strength and shape of any potential or
dressed gluon propagator) provided it supports the standard mechanism
of chiral  symmetry breaking \cite{BicudoRibeiro}, \cite{Orsay},
\cite{SteveandFelipe}.
symmetry properties of the $\Gamma$ pion vertex, are model dependent and
are of order $P^4$ (therefore vanishing at low energy). Lorentz invariance
restricts their form and allows only for two coefficients.

On a first glance, diagrams (\ref{defined1d2}, etc.) seem as difficult to calculate
as the full pion-pion scattering amplitude, but one needs to remember 
that  each $\Delta$ pion wavefunctions entering the
calculation can be taken to have only one power of the external $P$ to
this order, since there are four of them and they vanish by definition at
$P=0$, and the propagators can also be taken at $P=0$, leaving just one
momentum $k$ around the loop. This is a major simplification.

In the previous section we have shown how the numerical evaluation can be
performed with two very simple (too simple) kernels. Our results for $l_1$
and $l_2$ have phenomenologically correct signs and ratios, specially for the
gaussian toy model, see tables \ref{gaussianmodeltable} and \ref{rationalmodeltable} . 
Possibly because our results are computed at $P=0$ and 
without pion loops, our $l_1$ and $l_2$ seem too large in absolute value. 
This can be appreciated
\cite{Gasser,BCT,Yndurain,Pelaez,truong,Pich,Pennington,Espriu,NJL,Arriola,andrianov}
in tables \ref{lsfromas}, \ref{lsfromrho} and \ref{lstheo} which summarize the
present status of knowledge of these coefficients. The various conventions used
in the literature ($SU(3)$, $SU(2)$, renormalized, barred, etc.) have been unified
to the renormalized $l's$ at the scale of the $\rho$ meson. Some of those 
determinations carry information relative to finite quark (and therefore pion) masses,
to kaons and etas, or to pion loops, all of which have not been taken into account
in this work. The fairest comparison therefore is with the oldest results of Andrianov
\cite{andrianov}
in the large $N_c$ limit (no gluon corrections) and Pham and Truong
\cite{truong}. 
The later authors obtain the interesting relations
$$
l_2= \frac{f_\pi^2}{m_\rho^2}
$$
and
$$
l_1=\frac{1}{3} \frac{f_\pi^2}{m_\sigma^2} - 2 l_2 \ .
$$
If the $\sigma$ mass is sent to infinity, then the large $N_c$ ratio is recovered.
For finite $\sigma$ masses between 350 and 750 MeV we obtain the range given in table
\ref{lsfromrho}. 
This demonstrates the importance of repeating the model evaluations with kernels
whose meson excitations in various channels are known.

Our approach is potentially superior to the resonance saturation approximations since 
it includes the full vertex and ladder structures, that is, effects of continua and 
higher resonances. In particular we include the four pion direct interaction, and the 
exchange of the full series  of excited vector and scalar mesons. We also stress 
that the masses of our $\rho$ and $\sigma$ mesons are expected to be of the right order of 
magnitude, because 
our constituent quark mass have reasonable values of the order of $300 MeV$ to $400MeV$, 
see tables \ref{gaussianmodeltable} and \ref{rationalmodeltable}. 
Our approach is also potentially superior to Nambu-Jona-Lasinio determinations
by allowing to lift the approximation of contact interactions between quarks. Thus
we pave the way to calculating the $l_i$ coefficients from the lattice or from accurate
Schwinger-Dyson solutions. Finite current quark masses and meson loops remain to be 
incorporated to improve the precision of our calculation. We are currently considering 
some of these issues.

\begin{table}[h]%[H] add [H] placement to break table across pages
\caption{\label{lsfromas}
Phenomenological determinations of the l parameters (fits to scattering lengths or phase shifts 
in pion scattering.)}
\begin{ruledtabular}
\begin{tabular}{c|cc}
 Authors & $l_1^r(m_\rho) \cd 10^3$ & $l_2^r(m_\rho) \cd 10^3$ \\ \hline
Gasser \& Leutwyler & $-4.2 \pm 3.9$ & $9.0 \pm 2.7$ \\
Bijnens, Colangelo, Talavera; & & \\
Colangelo, Gasser, Leutwyler & $-2.2\pm 0.6$ & $9.0 \pm 2.7$ \\
Yndurain & $-4.1 \pm 0.7 $ & $9.8 \pm 0.5$ \\
G\'omez Nicola \& Pel\'aez & $-3.3 \pm 0.7$ & $4.8 \pm 0.6$
\end{tabular}
\end{ruledtabular}
\end{table}

\begin{table}[h]%[H] add [H] placement to break table across pages
\caption{\label{lsfromrho}
Phenomenological determinations of the l parameters 
(Based on $\rho$ meson resonance saturation).}
\begin{ruledtabular}
\begin{tabular}{c|cc}
 Authors & $l_1^r(m_\rho) \cd 10^3$ & $l_2^r(m_\rho) \cd 10^3$ \\ \hline
Gasser \& Leutwyler & -8.4 &  8.4 \\
Pham \& Truong & -5.5 to -24 & 14 \\
Ecker {\it et al.}& $-6.1 \pm 3.9$ & $5.3 \pm 2.7$ \\
Pennington \& Portol\'es& $-2.1 \pm 1.2$ & $5.9 \pm 1.0$
\end{tabular}
\end{ruledtabular}
\end{table}

\begin{table}[h]%[H] add [H] placement to break table across pages
\caption{\label{lstheo}
Theoretical determinations of the l parameters (based on the large $N_c$ 
approximation and/or the Nambu-Jona-Lasinio model).}
\begin{ruledtabular}
\begin{tabular}{c|cc}
 Authors & $l_1^r(m_\rho) \cd 10^3$ & $l_2^r(m_\rho) \cd 10^3$ \\ \hline
Espriu, de Rafael, Taron & -3.18 &  6.3 \\
Bijnens, Bruno, de Rafael & -4.8 & 6.4 \\
Ruiz Arriola & -7.4 & 7.8 \\
Andrianov & -3.2 & 6.4
\end{tabular}
\end{ruledtabular}
\end{table}

\acknowledgments
The authors acknowledge useful discussions with S. Cotanch, F. Kleefeld,
B. Hiller, P. Maris, J. R. Pelaez, A. Dobado, A. Gomez Nicola and specially E. Ribeiro.
Work partially supported by univ. Complutense on a travel grant, and by
grants  FPA 2000-0956, BFM 2002-01003 (Spain), 
F. J. Llanes-Estrada is thankful for the hospitality and scholarly
atmosphere at IST Lisbon.

\bibliography{apssamp}% Produces the bibliography via BibTeX.

%\vspace{ .6cm}

\appendix

%%%%%%%%%%%%%%%%%%%%%%%%%%%%%%%%%%%%%%%%%%%%%%%%%%%%%%%%%%%%%%%%%%
\section{Norm and $\protect f_\pi$}\label{normissues}
%%%%%%%%%%%%%%%%%%%%%%%%%%%%%%%%%%%%%%%%%%%%%%%%%%%%%%%%%%%%%%%%%%
In this appendix we present the proof of the well known fact 
that the normalization of the Bethe-Salpeter wavefunction is
the pion decay constant, rewritten in terms of our Ward Identity
techniques.
At null external pion momentum, $\chi^a(P=0)$ and $\Gamma_A^a$ are proportional
because eq. (\ref{GAvertex}) reduces to the homogeneous BS eq. (\ref{BS}).
Introduce an arbitrary proportionality constant $n_\pi$ by means of 
\be \label{chitogamma}
\chi^a(P=0)=\frac{\Gamma^a}{i n_\pi} \ ;
\ee
the $i$ guarantees $\chi$ to be real because of the explicit form of $\Gamma$ 
in eq. (\ref{GammaB}).

\par
The normalizing condition for the Bethe-Salpeter solution is, 
following Llewelyn-Smith \cite{LlewelynSmith},
\begin{eqnarray} \label{LSnormalization}
%
%
%>>>>>>>>>>>>>>>>>>>>>>>>>>>>>>>>>>>>>>>>>>>>>>>>>>>>>>>>>>>>>>
\begin{picture}(110,20)(0,0)
\put(0,0){
%\put(0,0){\framebox(110,20){}}
%
\put(0,0){
\begin{picture}(15,20)(0,0)
%%%%%%%%%%%%%%%%%%%%%%%%%%%% WF
\put(20,10){\oval(10,10)[l]}
\put(12,7.5){$\bullet$}
\put(-5,10){$\chi_{ _{-P}}^a$}
\end{picture}}
\put(20,0){
\begin{picture}(25,20)(0,0)
%%%%%%%%%%%%%%%%%%%%%%%%%%%%% parenthesis
\put(5,5 ){${\partial \over \partial_{P^\mu}}$}
\put(25,7){$\Bigl($}
\put(55,7){$\Bigr)$}
\end{picture}}
\put(45,0){
\begin{picture}(15,20)(0,0)
%%%%%%%%%%%%%%%%%%%%%%%%%%%%% propagators
\put(10,15){\line(1,0){5}}
\put(10,5){\vector(1,0){10}}
\put(25,15){\vector(-1,0){10}}
\put(25,5){\line(-1,0){5}}
\end{picture}}
\put(85,0){
\begin{picture}(10,20)(0,0)
%%%%%%%%%%%%%%%%%%%%%%%%%%%% WF
\put(3,7.5){$\bullet$}
\put(8,10){$\chi_{ _{P}}^b$}
\put(0,10){\oval(10,10)[r]}
\end{picture}}
}
\end{picture}
%<<<<<<<<<<<<<<<<<<<<<<<<<<<<<<<<<<<<<<<<<<<<<<<<<<<<<<<<<<<<<<
%
%
= 2 \, i \, P_\mu \delta^{ab}
\ .
\label{normfpi}
\end{eqnarray}
Because the right hand side of the normalization condition
is of first order in $P^\mu$ we can expand
in $\chi$ and in $\Gamma_A$ as in (\ref{expandGamma}).
There are two terms, depending on whether the derivative is applied
to the upper or lower propagator.

The first term can be written as
\begin{eqnarray} \label{totalnorm}
%>>>>>>>>>>>>>>>>>>>>>>>>>>>>>>>>>>>>>>>>>
\begin{picture}(75,40)(0,0)
\put(0,8){
%\put(0,-8){\framebox(75,40){}}
%%%%%%%%%%%%%%%%%%%%%%%%%%%%%% triangle
\put(20,0){\begin{picture}(40,40)(0,0)
\put(0,0){\line(1,0){30}}
\put(0,0){\vector(1,0){15}}
\put(30,0){\line(-1,1){15}}
\put(30,0){\vector(-1,1){10}}
\put(15,15){\line(-1,-1){15}}
\put(15,15){\vector(-1,-1){10}}
\put(-2,-2){$\bullet$}
\put(-20,-2){$\chi_{_{-P}}^a$}
\put(28,-2){$\bullet$}
\put(35,-2){$\chi_{_{P}}^b$}
\put(13,13){$\bullet$}
\put(15,20){$\partial_\mu {\cal S}^{-1}$}
\end{picture}}
}
\end{picture}
%<<<<<<<<<<<<<<<<<<<<<<<<<<<<<<<<
&=&
- {
%>>>>>>>>>>>>>>>>>>>>>>>>>>>>>>>>>>>>>>>>>
\begin{picture}(85,40)(0,0)
\put(10,8){
%\put(0,-8){\framebox(85,40){}}
%%%%%%%%%%%%%%%%%%%%%%%%%%%%%% triangle
\put(20,0){\begin{picture}(40,40)(0,0)
\put(0,0){\line(1,0){30}}
\put(0,0){\vector(1,0){15}}
\put(30,0){\line(-1,1){15}}
\put(30,0){\vector(-1,1){10}}
\put(15,15){\line(-1,-1){15}}
\put(15,15){\vector(-1,-1){10}}
\put(-2,-2){$\bullet$}
\put(-28,-2){${\Gamma_A}_{_{-P}}^a$}
\put(28,-2){$\bullet$}
\put(35,-2){${\Gamma_A}_{_{P}}^b$}
\put(13,13){$\bullet$}
\put(15,20){$\partial_\mu {\cal S}^{-1}$}
\end{picture}}
}
\end{picture}
%<<<<<<<<<<<<<<<<<<<<<<<<<<<<<<<<
\over
(i n_\pi )^2
}
\nonumber \\
+
{
%>>>>>>>>>>>>>>>>>>>>>>>>>>>>>>>>>>>>>>>>>
\begin{picture}(85,40)(0,0)
\put(10,8){
%\put(0,-8){\framebox(85,40){}}
%%%%%%%%%%%%%%%%%%%%%%%%%%%%%% triangle
\put(20,0){\begin{picture}(40,40)(0,0)
\put(0,0){\line(1,0){30}}
\put(0,0){\vector(1,0){15}}
\put(30,0){\line(-1,1){15}}
\put(30,0){\vector(-1,1){10}}
\put(15,15){\line(-1,-1){15}}
\put(15,15){\vector(-1,-1){10}}
\put(-2,-2){$\bullet$}
\put(-28,-2){${\Gamma_A}_{_{-P}}^a$}
\put(28,-2){$\bullet$}
\put(35,-2){$\chi_{_{P}}^b$}
\put(13,13){$\bullet$}
\put(15,20){$\partial_\mu {\cal S}^{-1}$}
\end{picture}}
}
\end{picture}
%<<<<<<<<<<<<<<<<<<<<<<<<<<<<<<<<
\over
i n_\pi
}
&+&
{
%>>>>>>>>>>>>>>>>>>>>>>>>>>>>>>>>>>>>>>>>>
\begin{picture}(75,40)(0,0)
\put(0,8){
%\put(0,-8){\framebox(75,40){}}
%%%%%%%%%%%%%%%%%%%%%%%%%%%%%% triangle
\put(20,0){\begin{picture}(40,40)(0,0)
\put(0,0){\line(1,0){30}}
\put(0,0){\vector(1,0){15}}
\put(30,0){\line(-1,1){15}}
\put(30,0){\vector(-1,1){10}}
\put(15,15){\line(-1,-1){15}}
\put(15,15){\vector(-1,-1){10}}
\put(-2,-2){$\bullet$}
\put(-20,-2){$\chi{_{-P}}^a$}
\put(28,-2){$\bullet$}
\put(35,-2){${\Gamma_A}_{_{P}}^b$}
\put(13,13){$\bullet$}
\put(15,20){$\partial_\mu {\cal S}^{-1}$}
\end{picture}}
}
\end{picture}
%<<<<<<<<<<<<<<<<<<<<<<<<<<<<<<<<
\over
i n_\pi
}
\label{separated loops}
\end{eqnarray}

The term with $\Gamma \Gamma$ can be shown to be zero. To see it,
one needs to take a derivative of eq. (\ref{SD}) which gives

\begin{equation} \label{DerivativeSD}
%>>>>>>>>>>>>>>>>>>>>>>>>>>>>>>>>>>>>>>>>>>>>>>>>>>
\begin{picture}(180,25)(0,0)
%\put(0,0){\framebox(180,25){}}
\put(0,10){$ S \, \partial_\mu S^{-1}   \,S \, \,=$}
\put(65,10){
%%%%%%%%%%%%%%%%%%%%%%%%%%%%%% half closed box
\begin{picture}(100,20)(0,7)
\put(0,15){\line(1,0){5}}
\put(0,5){\vector(1,0){10}}
\put(15,15){\vector(-1,0){10}}
\put(15,5){\line(-1,0){5}}
\put(15,0){\framebox(10,20){}}
\put(25,15){\line(1,0){5}}
\put(25,5){\vector(1,0){10}}
\put(40,15){\vector(-1,0){10}}
\put(40,5){\line(-1,0){5}}
\put(40,10){\oval(10,10)[r]}
\put(43,8){$\bullet$}
\put(49,10){$\partial_\mu S_0^{-1}$}
\end{picture}
}
\end{picture}
%<<<<<<<<<<<<<<<<<<<<<<<<<<<<<<<<<<<<<<<<<<<<<
\end{equation}
which applied to the $\Gamma \Gamma$ term in (\ref{totalnorm})
and employing eq. (\ref{upperWI2}) reduces it to

\begin{eqnarray}
%>>>>>>>>>>>>>>>>>>>>>>>>>>>>>>>>>>>>>>>>>
\begin{picture}(85,40)(0,0)
\put(10,8){
%\put(0,-8){\framebox(85,40){}}
%%%%%%%%%%%%%%%%%%%%%%%%%%%%%% triangle
\put(20,0){\begin{picture}(40,40)(0,0)
\put(0,0){\line(1,0){30}}
\put(0,0){\vector(1,0){15}}
\put(30,0){\line(-1,1){15}}
\put(30,0){\vector(-1,1){10}}
\put(15,15){\line(-1,-1){15}}
\put(15,15){\vector(-1,-1){10}}
\put(-2,-2){$\bullet$}
\put(-28,-2){$-{\Gamma_A}_{_{-P}}^a$}
\put(28,-2){$\bullet$}
\put(35,-2){${\Gamma_A}_{_{P}}^b$}
\put(13,13){$\bullet$}
\put(15,20){$\partial_\mu {\cal S}^{-1}$}
\end{picture}}
}
\end{picture}
%<<<<<<<<<<<<<<<<<<<<<<<<<<<<<<<<
&=&
%>>>>>>>>>>>>>>>>>>>>>>>>>>>>>>>>>>>>>>>>>>>>>>>>>>>
\begin{picture}(135,40)(0,0)
%\put(0,0){\framebox(135,40){}}
%%%%%%%%%%%%%%%%%%%%%%%%%%%%%% box with 3 vertices
%
\put(23,10){
\begin{picture}(40,20)(0,0)
%%%%%%%%%%%%%%%%%%%%%%%%%%%%%% box
\put(0,15){\line(1,0){5}}
\put(0,5){\vector(1,0){10}}
\put(15,15){\vector(-1,0){10}}
\put(15,5){\line(-1,0){5}}
\put(15,0){\framebox(10,20){}}
\put(25,15){\line(1,0){5}}
\put(25,5){\vector(1,0){10}}
\put(40,15){\vector(-1,0){10}}
\put(40,5){\line(-1,0){5}}
\end{picture}}
\put(63,10){
\begin{picture}(40,20)(0,0)
%%%%%%%%%%%%%%%%%%%%%%%%%%%%%% second box
\put(0,15){\line(1,0){5}}
\put(0,5){\vector(1,0){10}}
\put(15,15){\vector(-1,0){10}}
\put(15,5){\line(-1,0){5}}
\put(15,0){\framebox(10,20){}}
\put(25,15){\line(1,0){5}}
\put(25,5){\vector(1,0){10}}
\put(40,15){\vector(-1,0){10}}
\put(40,5){\line(-1,0){5}}
\end{picture}}
\put(-22,10){
\begin{picture}(40,20)(0,0)
%%%%%%%%%%%%%%%%%%%%%%%%%%%%%% half closed left
\put(45,10){\oval(10,10)[l]}
\put(37,8){$\bullet$}
\put(1,7){$-{\gamma_{\hspace{-.08cm}A}}_{_{P}} \frac{\sigma^a}{2}$}
\end{picture}}
\put(63,10){
\begin{picture}(40,20)(0,0)
%%%%%%%%%%%%%%%%%%%%%%%%%%%%%% half closed right
\put(40,10){\oval(10,10)[r]}
\put(43,8){$\bullet$}
\put(49,10){$\partial_\mu {\cal S}_0^{-1}$}
\end{picture}}
\put(23,10){
\begin{picture}(40,20)(0,0)
%%%%%%%%%%%%%%%%%%%%%%%%%%%%%% middle
\put(40,2){$\bullet$}
\put(40,13){$\bullet$}
\put(30,-6){$ S^{-1}$}
\put(30,21){$ {\Gamma_{\hspace{-.08cm}A}}_{_{-P}}^b$}
\end{picture}}
\end{picture}
%<<<<<<<<<<<<<<<<<<<<<<<<<<<<<<<<<<<<<<<<<<<<<<<<<<<
\nonumber \\
=
%>>>>>>>>>>>>>>>>>>>>>>>>>>>>>>>>>>>>>>>>>>>>>>>>>>>
\begin{picture}(90,20)(0,0)
%\put(0,0){\framebox(90,20){}}
%%%%%%%%%%%%%%%%%%%%%%%%%%%%%% loop with 2 vertices
%
\put(43,0){
\begin{picture}(40,20)(0,0)
%%%%%%%%%%%%%%%%%%%%%%%%%%%%%% lines
\put(0,15){\line(1,0){5}}
\put(0,5){\vector(1,0){10}}
\put(15,15){\vector(-1,0){10}}
\put(15,5){\line(-1,0){5}}
\end{picture}}
\put(18,0){
\begin{picture}(40,20)(0,0)
%%%%%%%%%%%%%%%%%%%%%%%%%%%%%% half closed left
\put(25,10){\oval(10,10)[l]}
\put(17,8){$\bullet$}
\put(-23,7){$-{\gamma_5}_{-P}{\gamma_{\hspace{-.08cm}A}}_{_{P}}$}
\end{picture}}
\put(58,0){
\begin{picture}(40,20)(0,0)
%%%%%%%%%%%%%%%%%%%%%%%%%%%%%% half closed right
\put(0,10){\oval(10,10)[r]}
\put(3,8){$\bullet$}
\put(9,10){$\partial_\mu {\cal S}^{-1}$}
\end{picture}}
\end{picture}
%<<<<<<<<<<<<<<<<<<<<<<<<<<<<<<<<<<<<<<<<<<<<<<<<<<<
&+&
%>>>>>>>>>>>>>>>>>>>>>>>>>>>>>>>>>>>>>>>>>>>>>>>>>>>
\begin{picture}(90,20)(0,0)
%\put(0,0){\framebox(90,20){}}
%%%%%%%%%%%%%%%%%%%%%%%%%%%%%% loop with 2 vertices
%
\put(25,0){
\begin{picture}(40,20)(0,0)
%%%%%%%%%%%%%%%%%%%%%%%%%%%%%% lines
\put(0,15){\line(1,0){5}}
\put(0,5){\vector(1,0){10}}
\put(15,15){\vector(-1,0){10}}
\put(15,5){\line(-1,0){5}}
\end{picture}}
\put(0,0){
\begin{picture}(40,20)(0,0)
%%%%%%%%%%%%%%%%%%%%%%%%%%%%%% half closed left
\put(25,10){\oval(10,10)[l]}
\put(17,8){$\bullet$}
\put(-2,7){$-{\Gamma_{\hspace{-.08cm}A}}_{_{P}}$}
\end{picture}}
\put(40,0){
\begin{picture}(40,20)(0,0)
%%%%%%%%%%%%%%%%%%%%%%%%%%%%%% half closed right
\put(0,10){\oval(10,10)[r]}
\put(3,8){$\bullet$}
\put(9,10){$ \partial_\mu {\cal S}_0^{-1}\ {\gamma_5}_{-P}$}
\end{picture}}
\end{picture}
%<<<<<<<<<<<<<<<<<<<<<<<<<<<<<<<<<<<<<<<<<<<<<<<<<<<
\end{eqnarray}
re-absorbing the remaining ladder, and employing the free propagator 
and bare axial vertex in eq. (\ref{barepropagator}, \ref{smallGAvertex}) 
this is proportional (because of the isospin factor not included) to 
\be
Tr \int (i \slash{P} S \partial_\mu S^{-1}S - \Gamma_{A_{-P}}S 
\frac{\gamma_\mu \gamma_5}{2i} S )\ .
\ee
Eliminating $\Gamma$ with its Ward Identity (\ref{GAvertex2}), and after some elementary
operations, this equals
\be \label{simplenormzero}
-i Tr \int (\gamma_\nu \partial_\mu S - \partial_\nu S \gamma_\mu) P_\nu = 0 
\ . \ee
Returning to equation (\ref{totalnorm}) and once the term with $\Gamma 
\Gamma$ has been shown to vanish, we need to evaluate the terms with a 
$\chi$ and a $\Gamma$.

\par
Diagrammatically again, one has (recall equations (\ref{upperWI2}),
 (\ref{Laddersaturation}) and (\ref{DerivativeSD}))
\begin{eqnarray}
&&
%>>>>>>>>>>>>>>>>>>>>>>>>>>>>>>>>>>>>>>>>>
\begin{picture}(85,40)(0,0)
\put(10,8){
%\put(0,-8){\framebox(85,40){}}
%%%%%%%%%%%%%%%%%%%%%%%%%%%%%% triangle
\put(20,0){\begin{picture}(40,40)(0,0)
\put(0,0){\line(1,0){30}}
\put(0,0){\vector(1,0){15}}
\put(30,0){\line(-1,1){15}}
\put(30,0){\vector(-1,1){10}}
\put(15,15){\line(-1,-1){15}}
\put(15,15){\vector(-1,-1){10}}
\put(-2,-2){$\bullet$}
\put(-28,-2){${\Gamma_A^a}_{_{-P}}$}
\put(28,-2){$\bullet$}
\put(35,-2){$\chi_{_{P}}^b$}
\put(13,13){$\bullet$}
\put(15,20){$\partial_\mu {\cal S}^{-1}$}
\end{picture}}
}
\end{picture}
%<<<<<<<<<<<<<<<<<<<<<<<<<<<<<<<<
\nonumber \\
&=&
%{P^2-M_\pi^2 \over i {\cal I}_P }
%>>>>>>>>>>>>>>>>>>>>>>>>>>>>>>>>>>>>>>>>>>>>>>>>>>>
\begin{picture}(135,40)(0,0)
%\put(0,0){\framebox(135,40){}}
%%%%%%%%%%%%%%%%%%%%%%%%%%%%%% box with 3 vertices
%
\put(23,10){
\begin{picture}(40,20)(0,0)
%%%%%%%%%%%%%%%%%%%%%%%%%%%%%% box
\put(0,15){\line(1,0){5}}
\put(0,5){\vector(1,0){10}}
\put(15,15){\vector(-1,0){10}}
\put(15,5){\line(-1,0){5}}
\put(15,0){\framebox(10,20){}}
\put(25,15){\line(1,0){5}}
\put(25,5){\vector(1,0){10}}
\put(40,15){\vector(-1,0){10}}
\put(40,5){\line(-1,0){5}}
\end{picture}}
\put(63,10){
\begin{picture}(40,20)(0,0)
%%%%%%%%%%%%%%%%%%%%%%%%%%%%%% second box
\put(0,15){\line(1,0){5}}
\put(0,5){\vector(1,0){10}}
\put(15,15){\vector(-1,0){10}}
\put(15,5){\line(-1,0){5}}
\put(15,0){\framebox(10,20){}}
\put(25,15){\line(1,0){5}}
\put(25,5){\vector(1,0){10}}
\put(40,15){\vector(-1,0){10}}
\put(40,5){\line(-1,0){5}}
\end{picture}}
\put(-22,10){
\begin{picture}(40,20)(0,0)
%%%%%%%%%%%%%%%%%%%%%%%%%%%%%% half closed left
\put(45,10){\oval(10,10)[l]}
\put(37,8){$\bullet$}
\put(20,7){$\frac{\chi_{_{P}}^b}{c}$}
\end{picture}}
\put(63,10){
\begin{picture}(40,20)(0,0)
%%%%%%%%%%%%%%%%%%%%%%%%%%%%%% half closed right
\put(40,10){\oval(10,10)[r]}
\put(43,8){$\bullet$}
\put(49,10){$\partial_\mu {\cal S}_0^{-1}$}
\end{picture}}
\put(23,10){
\begin{picture}(40,20)(0,0)
%%%%%%%%%%%%%%%%%%%%%%%%%%%%%% middle
\put(40,2){$\bullet$}
\put(40,13){$\bullet$}
\put(30,-6){$ S^{-1}$}
\put(30,21){$ {\Gamma^a_{\hspace{-.08cm}A}}_{_{-P}}$}
\end{picture}}
\end{picture}
%<<<<<<<<<<<<<<<<<<<<<<<<<<<<<<<<<<<<<<<<<<<<<<<<<<<
\nonumber \\
&=& 
%>>>>>>>>>>>>>>>>>>>>>>>>>>>>>>>>>>>>>>>>>>>>>>>>>>>
\begin{picture}(90,20)(0,0)
%\put(0,0){\framebox(90,20){}}
%%%%%%%%%%%%%%%%%%%%%%%%%%%%%% loop with 2 vertices
%
\put(43,0){
\begin{picture}(40,20)(0,0)
%%%%%%%%%%%%%%%%%%%%%%%%%%%%%% lines
\put(0,15){\line(1,0){5}}
\put(0,5){\vector(1,0){10}}
\put(15,15){\vector(-1,0){10}}
\put(15,5){\line(-1,0){5}}
\end{picture}}
\put(18,0){
\begin{picture}(40,20)(0,0)
%%%%%%%%%%%%%%%%%%%%%%%%%%%%%% half closed left
\put(25,10){\oval(10,10)[l]}
\put(17,8){$\bullet$}
\put(-23,7){${\gamma_5}\frac{\sigma^a}{2}
             \frac{\chi_{_{P}}^b}{c}$}
\end{picture}}
\put(58,0){
\begin{picture}(40,20)(0,0)
%%%%%%%%%%%%%%%%%%%%%%%%%%%%%% half closed right
\put(0,10){\oval(10,10)[r]}
\put(3,8){$\bullet$}
\put(9,10){$\partial_\mu {\cal S}^{-1}$}
\end{picture}}
\end{picture}
%<<<<<<<<<<<<<<<<<<<<<<<<<<<<<<<<<<<<<<<<<<<<<<<<<<<
\nonumber \\
&&+
%>>>>>>>>>>>>>>>>>>>>>>>>>>>>>>>>>>>>>>>>>>>>>>>>>>>
\begin{picture}(90,20)(0,0)
%\put(0,0){\framebox(90,20){}}
%%%%%%%%%%%%%%%%%%%%%%%%%%%%%% loop with 2 vertices
%
\put(25,0){
\begin{picture}(40,20)(0,0)
%%%%%%%%%%%%%%%%%%%%%%%%%%%%%% lines
\put(0,15){\line(1,0){5}}
\put(0,5){\vector(1,0){10}}
\put(-5,-5){$ ^{k+P/2} $}
\put(15,15){\vector(-1,0){10}}
\put(15,5){\line(-1,0){5}}
\end{picture}}
\put(0,0){
\begin{picture}(40,20)(0,0)
%%%%%%%%%%%%%%%%%%%%%%%%%%%%%% half closed left
\put(25,10){\oval(10,10)[l]}
\put(17,8){$\bullet$}
\put(-2,7){$\chi_{_{P}}^b$}
\end{picture}}
\put(40,0){
\begin{picture}(40,20)(0,0)
%%%%%%%%%%%%%%%%%%%%%%%%%%%%%% half closed right
\put(0,10){\oval(10,10)[r]}
\put(3,8){$\bullet$}
\put(9,10){$ \partial_\mu {\cal S}_0^{-1} {\gamma_5}_{-P} \frac{\sigma^a}{2}$}
\end{picture}}
\end{picture}
%<<<<<<<<<<<<<<<<<<<<<<<<<<<<<<<<<<<<<<<<<<<<<<<<<<<
\nonumber \\
&=& - \frac{1}{c}
%{P^2-M_\pi^2 \over i {\cal I}_P } 
Tr \{ {\gamma_5}_{-P} \frac{\sigma^a}{2}\chi_{_{P}^b} \,
\partial_\mu {\cal S} (k+P/2) \}
\nonumber \\
&&
+ { \frac{1}{2i}}
%>>>>>>>>>>>>>>>>>>>>>>>>>>>>>>>>>>>>>>>>>>>>>>>>>>>
\begin{picture}(90,20)(0,0)
%\put(0,0){\framebox(90,20){}}
%%%%%%%%%%%%%%%%%%%%%%%%%%%%%% loop with 2 vertices
%
\put(25,0){
\begin{picture}(40,20)(0,0)
%%%%%%%%%%%%%%%%%%%%%%%%%%%%%% lines
\put(0,15){\line(1,0){5}}
\put(0,5){\vector(1,0){10}}
\put(15,15){\vector(-1,0){10}}
\put(15,5){\line(-1,0){5}}
\end{picture}}
\put(0,0){
\begin{picture}(40,20)(0,0)
%%%%%%%%%%%%%%%%%%%%%%%%%%%%%% half closed left
\put(25,10){\oval(10,10)[l]}
\put(17,8){$\bullet$}
\put(-2,7){$\chi_{_{P}}^b$}
\end{picture}}
\put(40,0){
\begin{picture}(40,20)(0,0)
%%%%%%%%%%%%%%%%%%%%%%%%%%%%%% half closed right
\put(0,10){\oval(10,10)[r]}
\put(3,8){$\bullet$}
\put(9,10){$\gamma_\mu \gamma_5 \frac{\sigma^a}{2} $}
\end{picture}}
\end{picture}
%<<<<<<<<<<<<<<<<<<<<<<<<<<<<<<<<<<<<<<<<<<<<<<<<<<<
\ .
\end{eqnarray}
We get two terms. The first is of order $P^2$ because of the $c$ in the
denominator, and vanishes. 
The second is indeed non zero.
Going back now to eq. (\ref{LSnormalization}), we obtain
\be 
2 i P_\mu = \left( \frac{1}{i n_\pi} 2i \right) 
\begin{picture}(90,20)(0,0)
%\put(0,0){\framebox(90,20){}}
%%%%%%%%%%%%%%%%%%%%%%%%%%%%%% loop with 2 vertices
%
\put(25,0){
\begin{picture}(40,20)(0,0)
%%%%%%%%%%%%%%%%%%%%%%%%%%%%%% lines
\put(0,15){\line(1,0){5}}
\put(0,5){\vector(1,0){10}}
\put(15,15){\vector(-1,0){10}}
\put(15,5){\line(-1,0){5}}
\end{picture}}
\put(0,0){
\begin{picture}(40,20)(0,0)
%%%%%%%%%%%%%%%%%%%%%%%%%%%%%% half closed left
\put(25,10){\oval(10,10)[l]}
\put(17,8){$\bullet$}
\put(-2,7){$\chi_{_{P}}^b$}
\end{picture}}
\put(40,0){
\begin{picture}(40,20)(0,0)
%%%%%%%%%%%%%%%%%%%%%%%%%%%%%% half closed right
\put(0,10){\oval(10,10)[r]}
\put(3,8){$\bullet$}
\put(9,10){$\gamma_5 \gamma_\mu \frac{\sigma^a}{2} $}
\end{picture}}
\end{picture}
\ee
The definition of the weak decay constant, $f_\pi$ yields (with
no pion loops)
\be \label{deffpi}
\begin{picture}(90,20)(0,0)
%\put(0,0){\framebox(90,20){}}
%%%%%%%%%%%%%%%%%%%%%%%%%%%%%% loop with 2 vertices
%
\put(25,0){
\begin{picture}(40,20)(0,0)
%%%%%%%%%%%%%%%%%%%%%%%%%%%%%% lines
\put(0,15){\line(1,0){5}}
\put(0,5){\vector(1,0){10}}
\put(15,15){\vector(-1,0){10}}
\put(15,5){\line(-1,0){5}}
\end{picture}}
\put(0,0){
\begin{picture}(40,20)(0,0)
%%%%%%%%%%%%%%%%%%%%%%%%%%%%%% half closed left
\put(25,10){\oval(10,10)[l]}
\put(17,8){$\bullet$}
\put(-2,7){$\chi_{_{P}}^b$}
\end{picture}}
\put(40,0){
\begin{picture}(40,20)(0,0)
%%%%%%%%%%%%%%%%%%%%%%%%%%%%%% half closed right
\put(0,10){\oval(10,10)[r]}
\put(3,8){$\bullet$}
\put(9,10){$\gamma_5 \gamma_\mu \frac{\sigma^a}{2} $}
\end{picture}}
\end{picture}
=i P_\mu f_\pi \delta^{ab}
\ee
and therefore we must have
$$
n_\pi= f_\pi
$$
which immediately leads to equation (\ref{expandGamma}).
Finally, direct calculation of equation (\ref{deffpi}) leads to
\begin{widetext}
\be \label{finalfpi}
i f_\pi^2=3\int \frac{d^4q}{(2\pi)^4} \frac{1}{(A^2q^2-B^2)^2}
\left[E0\left(4AB+2q^2\left(B\frac{dA}{d(q^2)}-A\frac{dB}{d(q^2)}\right)
\right) +F0(2q^2A^2+4B^2)+q^2G0(B^2-q^2A^2)\right] \ .
\ee 
\end{widetext}
Notice the explicit color factor of $3$. All through the paper the
Bethe-Salpeter wavefunctions have been taken proportional to the identity
$\delta_{cc'}$ in color space. Had we normalized them in a different way,
say $\delta_{cc'}/\sqrt{3}$, this would immediately affect the formula
above reducing the factor to $\sqrt{3}$, the rest being absorbed by
the functions $E0$, $F0$, $G0$. The low energy theorems 
(Gell-Mann-Oakes-Renner theorem, Weinberg's amplitude, etc.) are unchanged
by this choice since the explicit form of these functions is never used to prove them:
they are always eliminated in terms of $f_\pi$. But the $l_1$, $l_2$
constants of the chiral lagrangian would indeed have to be rewritten in
terms of the modified Bethe-Salpeter wavefunctions. This of course would
not affect its numerical value. 

The normalizing condition (\ref{LSnormalization}),
\begin{eqnarray}
%
%
%>>>>>>>>>>>>>>>>>>>>>>>>>>>>>>>>>>>>>>>>>>>>>>>>>>>>>>>>>>>>>>
\begin{picture}(110,20)(0,0)
\put(0,0){
%\put(0,0){\framebox(110,20){}}
%
\put(0,0){
\begin{picture}(15,20)(0,0)
%%%%%%%%%%%%%%%%%%%%%%%%%%%% WF
\put(20,10){\oval(10,10)[l]}
\put(12,7.5){$\bullet$}
\put(-5,10){$\chi_{ _{-P}}$}
\end{picture}}
\put(20,0){
\begin{picture}(25,20)(0,0)
%%%%%%%%%%%%%%%%%%%%%%%%%%%%% parenthesis
\put(5,5 ){${\partial \over \partial_{P^\mu}}$}
\put(25,7){$\Bigl($}
\put(55,7){$\Bigr)$}
\end{picture}}
\put(45,0){
\begin{picture}(15,20)(0,0)
%%%%%%%%%%%%%%%%%%%%%%%%%%%%% propagators
\put(10,15){\line(1,0){5}}
\put(10,5){\vector(1,0){10}}
\put(25,15){\vector(-1,0){10}}
\put(25,5){\line(-1,0){5}}
\end{picture}}
\put(85,0){
\begin{picture}(10,20)(0,0)
%%%%%%%%%%%%%%%%%%%%%%%%%%%% WF
\put(3,7.5){$\bullet$}
\put(8,10){$\chi_{ _{P}}$}
\put(0,10){\oval(10,10)[r]}
\end{picture}}
}
\end{picture}
%<<<<<<<<<<<<<<<<<<<<<<<<<<<<<<<<<<<<<<<<<<<<<<<<<<<<<<<<<<<<<<
%
%
= 2 \, i \, P_\mu
\ .
\end{eqnarray}
which yields (\ref{deffpi}) can also be directly evaluated without using
Ward Identities.
By taking a derivative of this equation respect to $P^\mu$ (and contracting over $\mu$ as usual),
we get the following (derivatives respect to $P$ act only on the function immediately behind them and the
color factor is explicit).
\ba
8 i = 3 \int \frac{d^4 q}{(2 \pi)^4} Tr\left[ \right. \\ \nonumber
 2 \partial_\mu \chi_\pi(P) \partial^\mu S(q+P/2) \chi_\pi(-P) S(q-P/2) +   \\ \nonumber
2 \chi_\pi(P)\partial_\mu \partial^\mu S(q+P/2)\chi_\pi(-P) S(q-P/2)+ \\ \nonumber
2 \chi_\pi(P)\partial_\mu S(q+P/2) \partial^\mu \chi_\pi(-P)  S(q-P/2)+ \\ \nonumber
\left.
2  \chi_\pi(P) \partial^\mu S(q+P/2)\chi_\pi(-P)\partial_\mu S(q-P/2)+ 
\right] \\ \nonumber
\ea
To calculate this normalization one needs to make the wavefunction explicit
\be
\chi_\pi(P)= \gamma_5( E0(q^2) + F0(q^2) \slash{P} + G0(q^2) q\cd p \slash{q}+ \cdot \cdot \cdot) \,
\ee
which evaluated at $P=0$ yields
$$
\chi_\pi(0)=\gamma_5  E0(q^2) \ ; \ \partial_\mu^P \chi_\pi(P=0) = \gamma_5 (F0 \gamma_\mu + G0 \slash{q}
 q_\mu) \ .
$$
and make use of  the propagator expansion,  (\ref{prop0}, \ref{prop1}, \ref{prop2}) above.
A simple check on the resulting expression is to substitute $\chi$ by $\Gamma$, which yields
\begin{widetext}
$$
-3\int \frac{d^4q}{(2\pi)^4} \left[
-2B(\partial_\rho B) \partial^\rho \left( \frac{1}{A^2q^2-B^2}\right)-B^2 \partial_\rho
\partial^\rho  \left( \frac{1}{A^2q^2-B^2}\right)
\right]=0
$$

which vanishes upon employing Green's first identity. This checks the zero 
in (\ref{simplenormzero})
with a completely independent calculation, and is also approximately
observed in our computer codes. The result for $n_\pi$ is 
\ba \label{npiexplicit}
n_\pi^2 = 3i \int \frac{d^4q}{(2\pi)^4} \frac{1}{A^2 q^2-B^2}\left[
(AB'-BA') 2q^2 B (\ov{F}_0+q^2\ov{G}_0) \right. \\
\left. -AB^2(4\ov{F}_0+q^2\ov{G}_0 \right]
\ea
which, upon comparison with.
(\ref{finalfpi}) provides an integral constraint
between the Schwinger-Dyson and Bethe-Salpeter solutions. The barred
quantities have been defined in (\ref{Deltaexpansion}).
\end{widetext}

To conclude this discussion we recall the derivation of the
Gell-Mann-Oakes-Renner relation in the Bethe-Salpeter formalism.
this has been presented in \cite{weall}, \cite{Pedrosolo}, together
with the discussion of the Weinberg theorem.
At zero external pion momentum the Axial Ward Identity reads
\be
2im\Gamma^a_5(k,k)=\frac{\sigma^a}{2}(\gamma_5 S^{-1}+S^{-1}\gamma_5)
\ee
and as previously discussed
\be \label{explicitGamma5}
\Gamma^a_5=\frac{B_k}{m}\frac{\sigma^a}{2}\gamma_5
\ee
To start the simple demonstration of GMOR, ``undress'' the vertex 
$\Gamma^a_5$ to write
\begin{equation} 
%>>>>>>>>>>>>>>>>>>>>>>>>>>>>>>>>>>>>>>>>>>>>>>>>>>
\begin{picture}(180,25)(0,0)
%\put(0,0){\framebox(180,25){}}
\put(20,10){$S \, \Gamma_{5}^a \,S \, \,=$}
\put(65,10){
%%%%%%%%%%%%%%%%%%%%%%%%%%%%%% half closed box
\begin{picture}(100,20)(0,7)
\put(0,15){\line(1,0){5}}
\put(0,5){\vector(1,0){10}}
\put(15,15){\vector(-1,0){10}}
\put(15,5){\line(-1,0){5}}
\put(15,0){\framebox(10,20){}}
\put(25,15){\line(1,0){5}}
\put(25,5){\vector(1,0){10}}
\put(40,15){\vector(-1,0){10}}
\put(40,5){\line(-1,0){5}}
\put(40,10){\oval(10,10)[r]}
\put(43,8){$\bullet$}
\put(49,10){$\gamma_5\frac{\sigma^a}{2}$}
\end{picture}
}
\end{picture}
%<<<<<<<<<<<<<<<<<<<<<<<<<<<<<<<<<<<<<<<<<<<<<
\end{equation}
and, neglecting the contribution of higher pion states
(which exactly decouple in the chiral limit, as in ref. 
\cite{SteveandFelipe})
we can saturate the ladder by the pion pole yielding
\begin{equation}
\Gamma_5^a \simeq
\begin{picture}(90,20)(0,0)
%\put(0,0){\framebox(90,20){}}
%%%%%%%%%%%%%%%%%%%%%%%%%%%%%% loop with 2 vertices
%
\put(25,0){
\begin{picture}(40,20)(0,0)
%%%%%%%%%%%%%%%%%%%%%%%%%%%%%% lines
\put(0,15){\line(1,0){5}}
\put(0,5){\vector(1,0){10}}
\put(15,15){\vector(-1,0){10}}
\put(15,5){\line(-1,0){5}}
\end{picture}}
\put(0,0){
\begin{picture}(40,20)(0,0)
%%%%%%%%%%%%%%%%%%%%%%%%%%%%%% half closed left
\put(25,10){\oval(10,10)[l]}
\put(17,8){$\bullet$}
\put(-2,7){$\gamma_5 \frac{\sigma^a}{2}$}
\end{picture}}
\put(40,0){
\begin{picture}(40,20)(0,0)
%%%%%%%%%%%%%%%%%%%%%%%%%%%%%% half closed right
\put(0,10){\oval(10,10)[r]}
\put(3,8){$\bullet$}
\put(9,10){$\chi^a \frac{i}{P^2-M_\pi^2}\chi^a$}
\end{picture}}
\end{picture}
\end{equation}
substituting now eq. (\ref{chitogamma}) in the form
\be
\chi^a = \frac{S^{-1}\gamma^5+\gamma^5 S^{-1}}{i n_\pi}\frac{\sigma^a}{2}
\ee
and comparing with eq. (\ref{explicitGamma5}) we immediately obtain
$$
-2 m {\rm Tr} S = n_\pi^2 M_\pi^2
$$
corresponding to the GMOR relation upon identification of $n_\pi=f_\pi$,
consistent with the Llewelyn-Smith normalization condition.

We finally remind the reader of the explicit expression for the BCS
planar condensate:
\begin{eqnarray} \label{condensate}
\la \ov{\Psi}_u \Psi_u \ra = \la \ov{\Psi}_d \Psi_d \ra =
{\rm Tr} S = \\ \nonumber
i {\rm Tr} \int \frac{A_k \slash{k}+B_k}{A_k k^2-B_k^2  +i \varepsilon}
= \\ \nonumber
-3\int \frac{d^4k_E}{(2\pi)^4} \frac{4 B_k}{A_k^2 k^2 +B_k^2}
\end{eqnarray}

\end{document}